%% file: algoV3.tex
\documentclass[10pt]{article} 
\input{preamble}

\begin{document}

\title{Faster Provable Sieving Algorithms for the Shortest Vector Problem and the Closest Vector Problem on Lattices in $\ell_p$ Norm}

\author[1,2]{Priyanka Mukhopadhyay \thanks{mukhopadhyay.priyanka@gmail.com, p3mukhop@uwaterloo.ca}}

\affil[1]{Institute for Quantum Computing, University of Waterloo, Waterloo ON, Canada}
\affil[2]{Dept. of Combinatorics and Optimization, University of Waterloo, Waterloo ON, Canada}

\maketitle

\begin{abstract}
In this work, we give provable sieving algorithms for the Shortest Vector Problem (SVP) and the Closest Vector Problem (CVP) on lattices in $\ell_p$ norm ($1\leq p\leq\infty$). The running time we obtain is better than existing provable sieving algorithms. We give a new linear sieving procedure that works for all $\ell_p$ norm ($1\leq p\leq\infty$). The main idea is to divide the space into hypercubes such that each vector can be mapped efficiently to a sub-region. {We achieve a time complexity of $2^{2.751n+o(n)}$, which is much less than the $2^{3.849n+o(n)}$ complexity of the previous best algorithm.}
 We also introduce a mixed sieving procedure, where a point is mapped to a hypercube within a ball and then a quadratic sieve is performed within each hypercube. This improves the running time, especially in the $\ell_2$ norm, where we achieve a time complexity of $2^{2.25n+o(n)}$, while the List Sieve Birthday algorithm 
 has a running time of $2^{2.465n+o(n)}$.
 We adopt our sieving techniques to approximation algorithms for SVP and CVP in $\ell_p$ norm ($1\leq p\leq\infty$) and show that our algorithm has a running time of $2^{2.001n+o(n)}$, while previous algorithms have a time complexity of $2^{3.169n+o(n)}$.
 \end{abstract}

% Keywords
%\keyword{lattice; shortest vector problem; closest vector problem; provable sieving algorithm} 

% The fields PACS, MSC, and JEL may be left empty or commented out if not applicable
%\PACS{J0101}
%\MSC{}
%\JEL{}

%%%%%%%%%%%%%%%%%%%%%%%%%%%%%%%%%%%%%%%%%%
% Only for the journal Diversity
%\LSID{\url{http://}}

%%%%%%%%%%%%%%%%%%%%%%%%%%%%%%%%%%%%%%%%%%
% Only for the journal Applied Sciences:
%\featuredapplication{Authors are encouraged to provide a concise description of the specific application or a potential application of the work. This section is not mandatory.}
%%%%%%%%%%%%%%%%%%%%%%%%%%%%%%%%%%%%%%%%%%

%%%%%%%%%%%%%%%%%%%%%%%%%%%%%%%%%%%%%%%%%%
% Only for the journal Data:
%\dataset{DOI number or link to the deposited data set in cases where the data set is published or set to be published separately. If the data set is submitted and will be published as a supplement to this paper in the journal Data, this field will be filled by the editors of the journal. In this case, please make sure to submit the data set as a supplement when entering your manuscript into our manuscript editorial system.}

%\datasetlicense{license under which the data set is made available (CC0, CC-BY, CC-BY-SA, CC-BY-NC, etc.)}

%%%%%%%%%%%%%%%%%%%%%%%%%%%%%%%%%%%%%%%%%%

%\input{./intro}
\section{Introduction}
\label{sec:intro}

A lattice $\cL$ is the set of all integer combinations of linearly independent vectors $\bb_1,\dots,\bb_n \in \R^d$, 
\[
\cL = \cL(\bb_1, \ldots, \bb_n) := \{\sum_{i=1}^n z_i \bb_i : z_i \in \Z\} \;.
\]

 We call $n$ the rank of the lattice and $d$ the dimension of the lattice. 
 The matrix $\bB=(\bb_1,\dots,\bb_n)$ is called a basis of $\cL$. A lattice is said to be full-rank if $n=d$. 
 In this work, we only consider full-rank lattices unless otherwise stated. 

The two most important computational problems on lattices are the Shortest Vector Problem ({$\svp$}) %Please confirm whether this font needs to be kept. Reply : Yes. The font is fine.
and the Closest Vector 
Problem ($\cvp$). 
Given a basis for a lattice $\cL \subseteq \R^d$, the goal of $\SVP$ is to compute the shortest non-zero vector in $\cL$, 
while the goal of $\cvp$ is to compute a lattice vector at a minimum distance to a given target vector $\vect{t}$. 
Typically, the length/distance is defined in terms of the $\ell_p$ norm, which is given by
\begin{eqnarray}
\| \bx\|_p &=& (|x_1|^p + |x_2|^p + \cdots + |x_d|^p)^{1/p} \;\; \text{ for } \;\; 1 \le p < \infty \nonumber \\
\text{and } \|\bx\|_\infty &=& \max_{1 \le i \le d} |x_i| \nonumber
\end{eqnarray}
These lattice problems have been mostly studied in the Euclidean norm ($p=2$).
Starting with the seminal work of~\cite{1982_LLL}, algorithms for solving these problems either exactly or approximately have 
been studied intensely. These algorithms have found applications in various fields, such as
factoring polynomials over rationals~\cite{1982_LLL}, 
integer programming~\cite{1983_L,1987_K,2011_DPV,2011_EHN}, cryptanalysis~\cite{1990_O,1998_JS,2001_NS}, checking the solvability by radicals~\cite{1983_LM}, and solving low-density subset-sum problems~\cite{1992_CJLOSS}. 
More recently, many powerful cryptographic primitives have been constructed whose security is based on the {{worst-case} %Is the italics necessary? Reply : Not necessary. I have removed it.
} hardness of these or related lattice problems~\cite{1996_A,2007_MR,2009_G,2009_R,2011_BV,2013_BLPRS,2014_BV,Pei16,2017_DLLSSS}. 

\subsection{Prior Work}

The lattice algorithms that have been developed to solve $\SVP$ and $\CVP$ are either based on sieving techniques \cite{2001_AKS,2015_ADRS}, enumeration methods \cite{1985_FP,1987_K}, basis reduction \cite{1982_LLL,1987_S}, or Voronoi cell-based deterministic computation \cite{2013_MV,2011_DPV,2013_DV}. The fastest of these run in a time of $2^{c n}$, where $n$ is  the rank of the lattice and $c$ is some constant.
Since the aim of this paper is to improve time complexity of sieving algorithms, we mainly focus on these. For an overview of the other types of algorithms, interested readers can refer to the survey by Hanrot et al. \cite{2011_HPS}.

\subsubsection{Sieving Algorithms in the Euclidean Norm} 
\label{subsubsec:priorSieve2}

The first algorithm to solve $\SVP$ in the time exponential in the dimension of the lattice was given by  Ajtai, Kumar, and Sivakumar~\cite{2001_AKS} who devised a method based on ``randomized sieving'', whereby exponentially many randomly generated lattice vectors are iteratively combined to create increasingly short vectors, eventually resulting in the shortest vector in the lattice. The time complexity of this algorithm was shown to be $2^{3.4n+o(n)}$ by Micciancio and Voulgaris \cite{2010_MV}. This was later improved by Pujol and Stehle \cite{2009_PS}, who analyzed it with the birthday paradox and gave a time complexity of $2^{2.571n+o(n)}$. In \cite{2010_MV} the authors introduced List Sieve, which was modified in \cite{2009_PS} (List Sieve Birthday) to give a time complexity of $2^{2.465n+o(n)}$. The current fastest provable algorithm for exact SVP runs in a time of $2^{n+o(n)}$~\cite{2015_ADRS,2017_AS}, and the fastest algorithm that gives a large constant approximation runs in a time of $2^{0.802 n + o(n)}$~\cite{2011_LWXZ}. 

To make lattice sieving algorithms more practical for implementation, heuristic variants were introduced in \cite{2008_NV,2010_MV}. Efforts have been made to decrease the asymptotic time complexity at the cost of using more space \cite{2011_WLTB,2016_BDGL,2015_LW,2016_BL} and to study the trade-offs in reducing the space complexity \cite{2016_BL,2017_HK,2018_HKL,2018_LM}. Attempts have been made to make these algorithms competitive in high-performance computing environments \cite{2016_MB,2017_MLB,2017_YKYC,2018_D,2019_ADHKPS}. The theoretically fastest heuristic algorithm that is conjectured to solve $\svp$ runs in a time of $2^{0.29n+o(n)}$~\cite{2016_BDGL} (LDSieve).
%while in practice the Gauss Sieve appear to be the most practical in high dimensions \cite{2016_MB,2017_MLB,2017_YKYC}. %BDGL16,Laa15

The $\cvp$ is considered {to be} a harder problem than $\svp$ since there is a simple dimension and approximation-factor preserving 
reduction from $\svp$ to $\cvp$~\cite{1999_GMSS}. Based on a technique due to Kannan~\cite{1987_K}, Ajtai, Kumar, and Sivakumar~\cite{2002_AKS} gave a provable sieving based algorithm that gives a $1+\alpha$ approximation of $\CVP$ in time $(2+1/\alpha)^{O(n)}$. 
Later, exact exponential time algorithms for CVP were discovered~\cite{2013_MV,2015_ADS}. The current fastest algorithm for $\CVP$ runs in a time of $2^{n+o(n)}$ and is due to~\cite{2015_ADS}.

\subsubsection{Algorithms in Other $\ell_p$ Norms} Blomer and Naewe~\cite{2009_BN} and then Arvind and Joglekar~\cite{2008_AJ} generalized the AKS algorithm~\cite{2001_AKS} to give exact provable algorithms for $\SVP$ that run in a time of $2^{O(n)}$. 
%They in fact gave exact algorithm for a more general problem called the subspace avoidance problem $\sap$. 
%In particular, \cite{2009_BN} showed that several lattice problems, in particular the (approximate) $\SVP$ and the  (approximate) $\cvp$, are easily reducible to (approximate) $\sap$. 
Additionally,~\cite{2009_BN} gave a $1+\eps$ approximation algorithm for $\cvp$ for all $\ell_p$ norms that runs in a time of 
$(2+1/\eps)^{O(n)}$. For the special case when $p = \infty$, Eisenbrand et al.~\cite{2011_EHN} gave a $2^{O(n)} \cdot (\log (1/\eps))^n$ algorithm for $(1+\eps)$-approx CVP. Aggarwal and Mukhopadhyay \cite{2018_AM} gave an algorithm for $\svp$ and approximate $\cvp$ in the $\ell_{\infty}$ norm using a linear sieving technique that significantly improves the overall running time. In fact, for a large constant approximation factor, they achieved a running time of $3^n$ for $\SVP$. The authors have argued that it is not possible for any of the above-mentioned algorithms to achieve this running time in the $\ell_{\infty}$ norm. 

\subsubsection{Hardness Results} The first NP hardness result for $\CVP$ in all $\ell_p$ norms and $\SVP$ in the 
$\ell_\infty$ norm was given by Van Emde Boas~\cite{1981_vE}. Ajtai \cite{1998_A} proved that $\svp$ is $\NP$-hard under randomized reductions. Micciancio \cite{2001_M} showed that $\svp$ is $\NP$-hard to approximate within some constant approximation factor.
Subsequently, it was shown that approximating $\CVP$ in any $\ell_p$ norm and $\svp$ in $\ell_{\infty}$ norm up to a factor of $n^{c/\log \log n}$ is NP-hard~\cite{2003_DKRS,2000_D}. This difficulty of the approximation factor has been improved to $n^c$ in \cite{2019_M}, assuming the Projection Games Conjecture~\cite{2015_M}.
Furthermore, the difficulty of $\SVP$ up to factor $2^{\log^{1-\epsilon}n}$ has been obtained assuming $\NP\nsubseteq\RTIME(n^{\poly(\log n)})$ ~\cite{2005_K,2012_HR}. 
Recently,~\cite{2017_BGS} showed that for almost all $p \ge 1$, $\CVP$ in the $\ell_p$ norm cannot be solved in 
$2^{n (1-\eps)}$ of time under the strong exponential time hypothesis. 
A similar difficulty result has also been obtained for $\SVP$ in the $\ell_p$ norm \cite{2018_AsD}. 

%------------------------------------------------------------------------------------------
\subsection{Our Results and Techniques}

In this paper, we adopt the framework of \cite{2001_AKS,2002_AKS} and give sieving algorithms for $\svp$ and $\cvp$ in $\ell_p$
norm for $1\leq p\leq\infty$. The primary difference between our sieving algorithm and the previous AKS-style algorithms such as those in 
\cite{2001_AKS,2002_AKS,2009_BN,2008_AJ} is in the sieving procedure---ours is a linear sieve, while theirs is a quadratic sieve.
This results in an improvement in the overall \mbox{running time. }

Before describing our idea, we give an informal description of the sieving procedure of~\cite{2001_AKS,2002_AKS,2009_BN,2008_AJ}. 
The algorithm starts by randomly generating a set $S$ of $N = 2^{O(n)}$ lattice vectors with a length of at most $R = 2^{O(n)}$.
It then runs a sieving procedure a polynomial number of times. In the $i^{th}$ iteration, the algorithm starts with a list $S$ of lattice vectors of a length of at most $R_{i-1} \approx \gamma^{i-1} R$, for some parameter $\gamma \in (0,1)$. The algorithm maintains and updates a list of ``centers'' $C$, which is initialized to be the empty set.
%(The second vector in each centre pair is usually referred to as ``centre''.
Then, for each lattice vector $\vect{y}$ in the list, the algorithm checks whether there is a center $\vect{c}$ at a distance of at most $\gamma \cdot R_{i-1}$ from this vector. If there exists such a center, then the vector $\vect{y}$ is replaced in the list by $\vect{y} - \vect{c}$, and otherwise it is deleted from $S$ and added to $C$. This results in $N_{i-1} - |C|$ lattice vectors which have a length of at most $R_i \approx \gamma R_{i-1}$, where $N_{i-1}$ is the number of lattice vectors at the end of $i-1$ sieving iterations. We would like to mention here that this description hides many details and in particular, in order to show that this algorithm succeeds eventually in obtaining the shortest vector, we need to add a small perturbation to the lattice vectors to start with. The details of this can be found in Section~\ref{sec:aks_p}.

A crucial step in this algorithm is to find a vector $\vect{c}$ from the list of centers that is close to $\vect{y}$.  
This problem is called the nearest neighbor search (NNS) problem and has been well studied, especially in the context of 
heuristic algorithms for $\svp$ (see~\cite{2016_BDGL} and the references therein). 
A trivial bound on the running time for this is $|S| \cdot |C|$, but much effort has been dedicated to improving this bound under heuristic assumptions (see Section \ref{subsubsec:priorSieve2} for some references). Since they require heuristic assumptions, such improved algorithms for the NNS have not been used to improve the provable algorithms for $\svp$. 

One can also view such sieving procedures as a division of the ``ambient'' geometric space (consisting of all the vectors in the current list). In the {$i^{th}$} %We removed the superscript. Please confirm this revision. Reply : I think the superscript is required. So I have put it again.
iteration, the space of all vectors with a length of at most $R_{i-1}$ is divided into a number of sub-regions such that in each sub-region the vectors are within a distance of at most $\gamma R_{i-1}$ from a center. In the previous provable sieving algorithms such as those in \cite{2001_AKS,2009_BN,2008_AJ,2010_MV} or even the heuristic ones, these sub-regions have been an $\ell_p$ ball of certain radius (if the algorithm is in $\ell_p$ norm) or some sections of it (spherical cap, etc). Given a vector, one has to compare it with all the centers (and hence sub-regions formed so far) to determine in which of these sub-regions it belongs. If none is found, we make it a center and associate a new sub-region with it. {Note} that such a division of space depends on the order in which the vectors are processed.

\begin{figure}[h]
 \centering 
 \includegraphics[width=10cm, height=4cm]{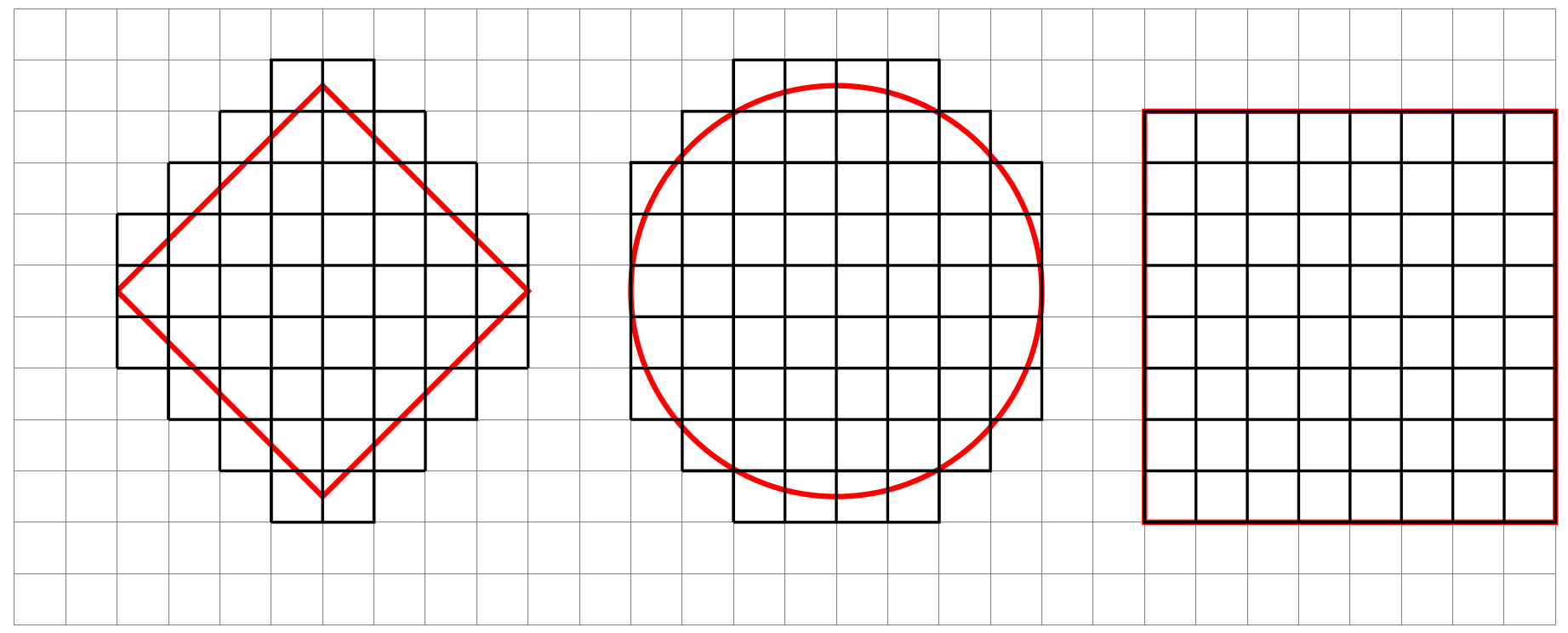}
 \caption[]{Division of the area of a circle in $\ell_1$, $\ell_2$, and $\ell_{\infty}$ norm (respectively) into smaller squares.}
  \label{fig:circle_grid}
\end{figure}

The basic idea behind our sieving procedure (let us call it Linear Sieve) is similar to that used in \cite{2018_AM,2018_AM2} in the special case of the $\ell_{\infty}$ norm. In fact, our procedure is a generalization of this method for all $\ell_p$ norm ($1\leq p\leq\infty$). We select these sub-regions as hypercubes and divide the ambient geometric space a priori (before we start processing the vectors in the current list) considering only the maximum length of a vector in the list. {A diagrammatic representation of such a division of space in two dimensions has been given in Figure \ref{fig:circle_grid}. It must be noted that in this figure (for ease of illustration), the radius of the small hypercube (square) is the same for $\ell_1, \ell_2$, and $\ell_{\infty}$ balls (circles). However, in our algorithm, this radius depends on the norm.} The advantage we obtain is that we can map a vector to a sub-region efficiently -in $O(n)$ time; i.e., in a sense we obtain better ``decodability'' property. If the vector's hypercube (sub-region) does not contain a center, we select this point as the center; otherwise, we subtract this vector from the center to obtain a shorter lattice vector. Thus, the time complexity of each sieving procedure is linear in the number of sampled vectors. {Overall, we obtain an improved time complexity at the cost of increased space complexity compared to previous \mbox{algorithms \cite{2008_AJ, 2009_BN, 2011_HPS}}. A more detailed explanation can be found in Section \ref{subsec:linSieve_p}. }

%\begin{comment}
%--------------------------------------------------------

Specifically, we obtain the \mbox{following result.}

\paragraph{{Theorem} %Please confirm whether this is a four-level heading or a list. Reply : Confirmed.
\ref{thm:multI_bday} in Section \ref{svpi:bday}}.

  Let $\gamma \in (0,1)$, and let $\xi >1/2$. 
  Given a full.rank lattice $\cL \subset \ratn^n$, there is a randomized algorithm for $\svpp$ with a success probability of at 
  least $1/2$, space complexity of at most $2^{\cspacep n+o(n)}$, and running time of at most $2^{\ctimep n+o(n)}$,
  where $\cspacep = \csp + \max(\ccp,\cbp/2)$\linebreak and $\ctimep = \max(\cspacep,\cbp)$, where 
  $\ccp =\log \left( 2+\frac{2}{\gamma}\right), \quad \csp = -\log\Big( 0.5 - \frac{1}{4\xi} \Big) $ and\linebreak
  $\cbp=\log \left(1 + \frac{2\xi(2-\gamma)}{1-\gamma} \right)$.

\subsection*{{A} mixed sieving algorithm}%Please confirm whether this is a secondary title or a list, if so, please modify.  Reply : This is a secondary title.

In an attempt to gain as many advantages as possible, we introduce a mixed sieving procedure (let us call it Mixed Sieve).
Here, we divide a hyperball into larger hypercubes so that we can map each point efficiently to a hypercube. Within a hypercube,
we perform a quadratic sieving procedure such as AKS with the vectors in that region. This improves both time and space complexity, especially in the Euclidean norm. 

\subsection*{{Approximation} algorithms for $\svpp$ and $\cvpp$}%Please confirm whether this is a secondary title or a list, if so, please modify.   Reply : This is a secondary title.

We have adopted our sieving techniques to approximation algorithms for $\svpp$ and $\cvpp$. The idea is quite similar to
that described in \cite{2018_AM,2018_AM2} (where it was shown to work for only the $\ell_{\infty}$ norm). 
In Section \ref{sec:svpi-approx}, we have shown that our approximation algorithms are faster than those of \cite{2008_AJ,2009_BN}, but again they require more space.

\begin{Remark}
It is quite straightforward to extend our algorithm to the Subspace Avoiding Problem ($\SAP$) (or Generalized Shortest Vector Problem $\gsvp$) \cite{2009_BN,2008_AJ}: replace the quadratic sieve by any one of the faster sieves described in this paper. We thus obtain a similar improvement in running time. By Theorem 3.4 in \cite{2009_BN}, there are polynomial time reductions from other lattice problems such as the Successive Minima Problem ($\smp$) {({given} a lattice $\cL$ with rank $n$, the Successive Minima Problem ($\smp$) requires to find $n$ linearly independent vectors $\vect{v}_1,\ldots,\vect{v}_n\in\cL$ such that $\|\vect{v}_i\|_p\leq c\lambda_i^{(p)}(\cL)$.)} 
and Shortest Independent Vector Problem ($\sivp$) {({given} a rank $n$ lattice $\cL$ the Shortest Independent Vector Problem ($\sivp$) requires to find $n$ linearly independent vectors $\vect{v}_1,\ldots\vect{v}_n\in\cL$ such that $\|\vect{v}_i\|_p\leq c\lambda_n^{(p)}(\cL)$.
The definition of $\minp_i$ (and hence $\minp_n$) has been given in Section \ref{sec:prelim} (Definition \ref{defn:succMin}); $c$ is the approximation factor)} 
with approximation factor $1+\epsilon$ to $\gsvp$ with approximation factor $1+\epsilon$. Thus, we can obtain a similar improvement in running time for both these problems. Since in this paper, we focus mainly on $\SVP$ and $\CVP$, we do not delve into further details for these other problems.
 \label{remark:sap}
\end{Remark}

\begin{Remark}

Our algorithm (and in that case any sieving algorithm) is quite different from deterministic algorithms such as those in \cite{2011_DPV,2013_DK}. %These are derandomization of AKS sieving algorithm. 
 They reduce the problem in any norm to a $\ell_2$ norm and compute an approximation of the shortest vector length (or distance of the closest lattice point to a target in case of $\CVP$) using the Voronoi cell-based deterministic algorithm in \cite{2010_MV}. Then, they enumerate all lattice points within a convex region to find the shortest one. Constructing ellipsoidal coverings, it has been shown that the lattice points within a convex body can be computed in a time proportional to the maximum number of lattice points that the body can contain in any translation of an ellipsoid. Note for $\ell_p$ norm that any smaller $\ell_q$ ball (where $p=q$ or $p\neq q$) can serve this purpose, and the bound on the number of translates comes from standard packing arguments. For these deterministic algorithms, the target would be to chose a shape so that the upper bound (packing bound) on the number of translates can be reduced. Thus, the authors chose small $\ell_p$ balls to cover a larger $\ell_p$ ball.

In contrast, in our sieving algorithm, we aimed to map each lattice point efficiently within a sub-region. Thus, we divided any arbitrary $\ell_p$ ball into smaller hypercubes. The result was an increase in space complexity, but due to the efficient mapping, we reduced the running time. To the best of our knowledge, this kind of sub-divisions has not been used before in any sieving algorithm. The focus of our paper is to develop randomized sieving algorithms. Thus, we will not delve further into the details of the above-mentioned deterministic algorithms. Clearly, these are different procedures.

 \label{remark:enum}
\end{Remark}

%------------------
\begin{comment}
{\subsection{Future Work}

An obvious direction for further research would be to design heuristic algorithms on these kind of sieving techniques and to study if these can be adapted to other computing environments, such as parallel computing.   

The major difference between our algorithm and others such as \cite{2001_AKS,2009_BN} is in the choice of the shape of the sub-regions in which we divide the ambient space (as has already been explained before). Due to this, we obtain superior ``decodability'' in the sense that a vector can be efficiently mapped to a sub-region at the cost of inferior space complexity, as described before. It might be interesting to study what other shapes of these sub-regions might be considered and what trade-offs are obtained. 

It might be possible to improve the bound on the number of hypercubes required to cover the hyperball. At least in the $\ell_{\infty}$ norm, we have seen that the number of hypercubes may depend on the initial position of the smaller hypercube, whose translates cover the larger hyperball. In fact, it might be possible to obtain some lower bound on the complexity of this kind of approach.}
\end{comment}
%-----------------------------
\subsection{Organization of the Paper}

In Section \ref{sec:prelim}, we give some preliminary definitions and results that are useful for this paper. In Section \ref{sec:aks_p}, we introduce the linear sieving technique, while in Section \ref{sec:subQuad}, we describe the mixed sieving technique. In Section \ref{sec:approx}, we discuss how to extend our sieving methods to approximation algorithms.

\section{Preliminaries}
\label{sec:prelim}

\subsection{Notations}
We write $\log_q$ to represent the logarithm to the base $q$, and simply $\log$ when the base is $q=2$.
We denote the natural logarithm by $\ln$.

We use bold lowercase letters (e.g., $\vect{v}^n$) for vectors and bold uppercase letters for matrices 
(e.g., $\vect{M}^{m\times n}$).
We may drop the dimension in the superscript whenever it is clear from the context.
Sometimes, we represent a matrix as a vector of column (vectors) 
(e.g., $\vect{M}^{m\times n} = [\vect{m}_1 \vect{m}_2 \ldots \vect{m}_n] $ where each $\vect{m}_i$ is an $m-$length vector).
The $i^{th}$ co-ordinate of $\vect{v}$ is denoted by $v_i$.
%For any set of vectors $\vect{S} = \{ \vect{s}_1, \ldots \vect{s}_n \}$ let $\|\vect{S}\|_p = \max_{i=1}^n ||\vect{s}_i||_p$.
%Unless mentioned otherwise, throughout this paper we will be working in $\real^n$.
%The dimension may vary and will be specified.
%We denote points in $\real^n$ by capital letters and sometimes we may mention the vectors corresponding to those points in
%brackets. For example if $\vect{x}$ is the vector corresponding to point $X$, we denote it as $X(\vect{x})$.

Given a vector $\vect{x} = \sum_{i=1}^n x_i \vect{m}_i$ with $x_i \in \ratn$, the representation size of $\vect{x}$ with
respect to $\vect{M}$ is the maximum of $n$ and the binary lengths of the numerators and denominators of the 
coefficients $x_i$.

We denote the volume of a geometric body $A$ by $\vol(A)$.

%$|A|$ denotes volume of A if it is a geometric body and cardinality if it is a set.

%---------------------------------------------------------------

\subsection{$\ell_p$ Norm and Ball}

\begin{Definition}

The \textbf{$\ell_p$ norm} of a vector $\vect{v} \in \real^n$ is defined by\\
$\|\vect{v}\|_p = \Big(\sum_{i=1}^n |v_i|^p \Big)^{1/p}$
for $1 \leq p < \infty$ and $\|\vect{v}\|_{\infty} = \max \{ |v_i| : i=1, \ldots n \}$ for $p=\infty$.
 
\end{Definition}

\begin{fact}
For $\vect{x} \in \real^n \quad \|\vect{x}\|_p \leq \|\vect{x}\|_2 \leq \sqrt{n} \|\vect{x}\|_p$ for $p \geq 2$ and \\
$\frac{1}{\sqrt{n}} \|\vect{x}\|_p \leq \|\vect{x}\|_2 \leq \|\vect{x}\|_p$ for $1 \leq p < 2$.

 \label{fact:lp}
\end{fact}

\begin{Definition}

A \textbf{ball} is the set of all points within a fixed distance or radius (defined by a metric) from a fixed point or center.
More precisely, we define the (closed) ball centered at $\vect{x} \in \real^n$ with radius $r$ as
$$ \ballp_n (\vect{x},r) = \{ \vect{y} \in \real^n : \|\vect{y}-\vect{x}\|_p \leq r \}$$.
\end{Definition}
The boundary of $\ballp_n (\vect{x},r) $ is the set 
\begin{equation}\bd(\ballp_n (\vect{x},r)) = \{ \vect{y} \in \real^n : \|\vect{y}-\vect{x}\|_p = r \}.\nonumber\end{equation}

We may drop the first argument when the ball is centered at the origin $\vect{0}$ and drop both arguments for a unit ball 
centered at the origin.
Let\\ $\ballp_n (\vect{x},r_1,r_2) = \ballp_n (\vect{x},r_2) \setminus \ballp_n (\vect{x},r_1) = 
\{ \vect{y} \in \real^n : r_1 < \|\vect{y}-\vect{x}\|_p \leq r_2 \}$.
We drop the first argument if the spherical shell or corona is centered at the origin.

%--------
\begin{fact}
 $|\ballp_n (\vect{x},c\cdot r)| = c^n\cdot|\ballp_n (\vect{x},r)|$ for all $c > 0$.
 
 \label{fact:ballRad}
\end{fact}
%---------
\begin{fact}
 $\vol(\ballp_n(R)) = \frac{\Big(2\Gamma\left(\frac{1}{p}+1\right)R\Big)^n}{\Gamma\left(\frac{n}{p}+1\right)}$.
 Specifically $\vol(\balli_n(R)) = (2R)^n$.
 
 \label{fact:ballVol}
\end{fact}

The algorithm of Dyer, Frieze, and Kannan \cite{1991_DFK}  almost uniformly selects a point in any convex body in polynomial
time if a membership oracle is given \cite{2000_GG}.
For the sake of simplicity, we ignore the implementation detail and assume that we are able to uniformly select a point
in $\ballp_n (\vect{x},r)$ in polynomial time.

%---------------------------------------------------------------

\subsection{Lattice}

\begin{Definition}
 A \textbf{lattice} $\cL$ is a discrete additive subgroup of $\real^{d}$.
 Each lattice has a basis \linebreak$\vect{B} = [\vect{b}_1, \vect{b}_2, \ldots \vect{b}_n]$, where $\vect{b}_i \in \real^{d}$ and
 \begin{eqnarray}
  \cL=\cL(\vect{B}) = \Big\{ \sum_{i=1}^n x_i\vect{b}_i : x_i \in \intg \quad \text{ for } \quad 1 \leq i \leq n\Big\}
  \nonumber
 \end{eqnarray}
\end{Definition}
For algorithmic purposes, we can assume that $\cL \subseteq \ratn^{d}$.
We call $n$ the \emph{rank} of $\cL$ and $d$ the \emph{dimension}.
If $d=n$, the lattice is said to be full-rank.
Though our results can be generalized to arbitrary lattices, in the rest of the paper, we only consider full-rank lattices.

\begin{Definition}
 For any lattice basis $\vect{B}$, we define the \textbf{{fundamental} %Is the bold necessary?  Reply : It is good to have it bold, since that is the convention followed for definitions throughout the paper.
 parallelepiped} as
 \begin{eqnarray}
  \fpar(\vect{B}) = \{ \vect{Bx} : \vect{x} \in [0,1)^n \}	\nonumber
 \end{eqnarray}
\end{Definition}
If $\vect{y} \in \fpar(\vect{B})$, then $\|\vect{y}\|_p \leq n\|\vect{B}\|_p $, as can be easily seen by triangle inequality.
For any $\vect{z} \in \real^{n}$, there exists a unique $\vect{y} \in \fpar(\vect{B})$ such that 
$\vect{z}-\vect{y} \in \cL(\vect{B})$.
This vector is denoted by $\vect{y} \equiv \vect{z} \mod \vect{B}$ and it can be computed in polynomial time given $\vect{B}$
and $\vect{z}$.

\begin{Definition}
 For $i \in [n]$, the \textbf{$i^{th}$ successive minimum} is defined as the smallest real number $r$ such that $\cL$ contains $i$ 
 linearly independent vectors with a length of at most $r$:
 \begin{eqnarray}
  \minp_i (\cL) = \inf \{ r : \dim( \Span(\cL \cap \ballp_n (r)) ) \geq i \}	\nonumber
 \end{eqnarray}
 \label{defn:succMin}
\end{Definition}
Thus, the \emph{first successive minimum} of a lattice is the length of the shortest non-zero vector in the lattice:
\begin{eqnarray}
 \minp_1 (\cL) = \min \{ \|\vect{v}\|_p : \vect{v} \in \cL \setminus \{\vect{0}\} \}	\nonumber
\end{eqnarray}

%-----------------------
We consider the following lattice problems.
In all the problems defined below, $c \geq 1$ is some arbitrary approximation factor (usually specified as subscript), 
which can be a constant or a function of any parameter of the lattice (usually rank).
For exact versions of the problems (i.e., $c=1$), we drop the subscript.

\begin{Definition}[\textbf{{Shortest} %Is the bold necessary?  Reply : As said earlier, it is good to have it bold.
 Vector Problem ($\svp_c^{(p)}$)}]

Given a lattice $\cL$, find a vector $\vect{v} \in \cL \setminus \{\vect{0}\}$ such that
$\|\vect{v}\|_p \leq c \|\vect{u}\|_p$ for any other $\vect{u} \in \cL \setminus \{\vect{0}\}$.

%Let $\minp_M(L) = \min \{ r \in \real : \exists \vect{v} \in L \setminus M, \|\vect{v}\|_p \leq r \}$ is the subspace avoiding minimum.

%The \textbf{Shortest Vector Problem ($\svp_c^{(p)}$)} is a special instance of $\sap_c^{(p)}$ when $M = \{ \vect{0} \}$.

\end{Definition}

%-------------------------

\begin{Definition}[\textbf{Closest Vector Problem ($\cvp_c^{(p)}$)}]

Given a lattice $\cL$ with rank $n$ and a target vector $\vect{t} \in \real^n$, find $\vect{v} \in \cL$ such that
$\|\vect{v}-\vect{t}\|_p \leq c \|\vect{w}-\vect{t}\|_p$ for all other $\vect{w} \in \cL$.
 
\end{Definition}

%\begin{definition}[\textbf{$\gapcvp_c^{(p)}$}]
 
 %Given a rank $n$ lattice $L$, a target vector $\vect{t}$ and $\alpha \in \real_{>0}$, either find $\vect{v} \in L$ with $\|\vect{v}-\vect{t}\|_p \leq \alpha$ or state that $\|\vect{w}-\vect{t}\|_p > \frac{\alpha}{c}$ for all   $\vect{w} \in L$.
 
%\end{definition}

 %---------------
 \begin{Lemma}[\cite{2018_AM2}]

The LLL algorithm \cite{1982_LLL} can be used to solve ${\svp_{2^n}^{(p)}}$ in polynomial time.
 
 \label{lem:LLL}
\end{Lemma}

The following result shows that in order to solve $\svp_{1+\epsilon}^{(p)}$, it is sufficient to consider the case when 
$2 \leq \minp_1(\cL) < 3$. This is done by appropriately scaling the lattice. 

\begin{Lemma}[\textbf{Lemma 4.1 in } \cite{2009_BN}]

For all $\ell_p$ norms, if there is an algorithm $A$ that for all lattices $\cL$ with $2 \leq \minp_1 (\cL) < 3 $
 solves $\svp_{1+\epsilon}^{(p)}$ in time $T=T(n,b,\epsilon)$, then there is an
algorithm $A'$ that solves $\svp_{1+\epsilon}^{(p)}$ for all lattices in time $O(nT+n^4b)$.

\end{Lemma}
Thus, henceforth, we assume $2 \leq \minp_1(\cL) < 3$.

%---------------------------------------------------------------------------------------------------------
%	BALL INTERSECT
%----------------------------------------------------------------------------------------------------------
\subsection{Some Useful Definitions and Results}
\label{subsec:vol}

In this section, we give some results and definitions which are useful for our analysis later.

\begin{Definition}
 Let $P$ and $Q$ are two point sets in $\real^n$. The \textbf{Minkowski sum} of $P$ and $Q$, denoted as $P\oplus Q$, is the point set $\{p+q : p\in P, q\in Q\}$.
\end{Definition}

%Let $\kappa_n$ be the volume of the unit ball $\ballp_n$.

\begin{Lemma}
 
 Let $B_1=\ballp_n(\vect{0},a)$ and $B_2=\ballp_n(\vect{v},a)$ such that $\|\vect{v}\|_p=\minp_1$ and $\minp_1<2a$.
 Let $D = B_1 \cap B_2$.
 
 If $|D|$ and $|B_1|$ are the volumes of $D$ and $B_1$, respectively, then
 \begin{enumerate}
  \item \cite{2007_BN} $\frac{|D|}{|B_1|} \geq 2^{-n} \Big( 1 -  \frac{\minp_1}{2a} \Big)^n \text{ if } 1 \leq p < \infty. $
  \item \cite{2011_HPS} When $p=2$, further optimization can be done such that we get\\
  $\frac{|D|}{|B_1|} \geq \Big[1-\Big(\frac{\mint_1}{2a}\Big)^2\Big]^{n/2}$.
  \item \cite{2018_AM} When $p=\infty$ then $\frac{|D|}{|B_1|} \geq \Big(1-\frac{\mini_1}{2a} \Big)^n$.
 \end{enumerate}
 
 \label{lem:overlap}
\end{Lemma}

%--------------
\begin{Theorem}[\textbf{Kabatiansky and Levenshtein} \cite{1978_KL}]
 Let $E\subseteq\real^n\setminus\{\vect{0}\}$. If there exists $\phi_0>0$ such that for any $\vect{u},\vect{v}\in E$, we have
 $\phi_{\vect{u},\vect{v}}\geq\phi_0$, then $|E|\leq 2^{cn+o(n)}$ with 
 $c=-\frac{1}{2}\log[1-\cos(\min(\phi_0,62.99^{\circ}))]-0.099$.
 
 {Here, $\phi_{\vect{u},\vect{v}}$ is the angle between the vectors $\vect{u}$ and $\vect{v}$.}
 \label{thm:KL}
\end{Theorem}

Below, we give some bounds which work for all $\ell_p$ norms. We especially mention the bounds obtained for the $\ell_2$ norm where 
some optimization has been performed using Theorem \ref{thm:KL}.
\begin{Lemma}
\begin{enumerate}
 \item \cite{2009_BN} Let $\ccp = \log(1+\frac{2}{\gamma})$. If $\centrep$ is a set of points in $\ballp_n(R)$ such that the distance between two 
 points is at least $\gamma R$, then $|\centrep| \leq 2^{\ccp n+o(n)}$.
 \item \cite{2010_MV,2011_HPS} When $p=2$, we can have $|\centret| \leq 2^{\cc n+o(n)}$ where $\cc=-\log\gamma+0.401$. 
\end{enumerate}
 \label{lem:cc2}
\end{Lemma}
Since the distance between two lattice vectors is at most $\minp_1(\cL)$, we obtain the following corollary.
\begin{Corollary}
Let $\cL$ be a lattice and $R$ be a real number greater than the length of the shortest vector in the lattice.
 \begin{enumerate}
  \item \cite{2007_BN} $|\ballp_n(R)\cap\cL|\leq 2^{\cbp n}$ where $\cbp=\log\Big(1+\frac{2R}{\minp_1}  \Big) $. 
  \item \cite{2009_PS,2011_HPS} $|\ballt_n(R)\bigcap\cL|\leq 2^{\cb n+o(n)}$ where $\cb=\log\frac{R}{\mint_1}+0.401$.
 \end{enumerate}
\label{cor:latBall}
\end{Corollary}

\vspace{-12pt}

\section{A Faster Provable Sieving Algorithm in $\ell_p$ Norm}
\label{sec:aks_p}

In this section, we present an algorithm for $\svpp$ that uses the framework of the AKS algorithm \cite{2001_AKS} but uses
a different sieving procedure that yields a faster running time. 
Using Lemma~\ref{lem:LLL}, we can obtain an estimate 
$\lambda^*$ of $\minp_1(\cL)$ such that $\minp_1(\cL) \le \lambda^* \le 2^n \cdot \minp_1(\cL)$. 
Thus, if we try polynomially many different values of $\lambda = (1+1/n)^{-i} \lambda^*$, for $i \ge 0$, then for one of them,
we have $\minp_1(\cL) \le \lambda \le (1+1/n) \cdot \minp_1(\cL)$.
For the rest of this section, we assume that we know an estimated $\lambda$ of the length of the shortest vector in $\cL$, which 
is correct up to a factor $1 + 1/n$. 

The AKS algorithm (or its $\ell_p$ norm generalization in \cite{2008_AJ,2009_BN}) initially uniformly samples  a large number of perturbation vectors, $\vect{e} \in \ballp_n(d)$, where $d \in \real_{>0}$,
and for each such perturbation vector, it maintains a vector $\vect{y}$ close to the lattice ($\vect{y}$ is such that 
$\vect{y}-\vect{e} \in \cL$).
%This is done due to difficulty in analysis caused by distributions over $\cL \cap \balli_n(R)$, where $R \in \real_{>0}$.
Thus, initially, we have a set $S$ of many such pairs $(\vect{e},\vect{y}) \in \ballp_n(d) \times \ballp_n(R)$ for some $R \in 2^{O(n)}$. 
The desired situation is that after a polynomial number of such sieving iterations, we are left with a set of vector pairs
$(\vect{e}'',\vect{y}'')$ such that $\vect{y}''-\vect{e}'' \in \cL \cap \ballp_n(O(\minp_1(\cL)))$. 
Finally, we take the pair-wise differences of the lattice vectors corresponding to these vector pairs and output the one
with the smallest non-zero norm. 
It was shown in~\cite{2001_AKS,2008_AJ,2009_BN} that, with overwhelming probability, this is the shortest vector in the lattice. 

\begin{figure}[h]
\centering
\includegraphics[width=5cm, height=4cm]{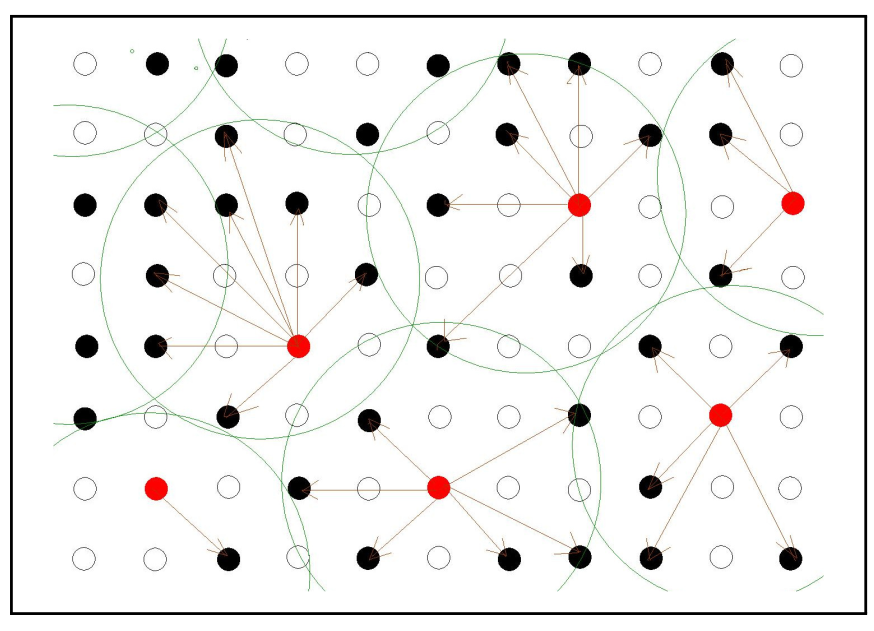}
 \caption{{One iteration of the quadratic AKS sieve in the $\ell_2$ norm. Each point represents a vector pair. The solid dots are the sampled ones, while the hollow dots are the unsampled ones. Among the sampled vector pairs, some are identified as centers (red dots) and the space is divided into a number of balls, centered around these red dots. Vector subtraction (denoted by arrow) is performed with the center pair in each ball, such that we obtain shorter lattice vectors in the next iteration.  }}
 \label{fig:aks}
\end{figure}

One of the main and usually the most expensive steps in this algorithm is the sieving procedure, where given a list of vector 
pairs $(\vect{e},\vect{y}) \in \ballp_n(d) \times \ballp_n(R)$ in each iteration, it outputs a list of vector pairs 
$(\vect{e}',\vect{y}') \in \ballp_n(d) \times \ballp_n(\gamma R)$ where $\gamma \in \real_{(0,1)}$. 
In each sieving iteration, a number of vector pairs (usually exponential in $n$) are identified as ``center pairs''.
The second element of each such center pair is referred to as the ``center''.
By a well-defined map, each of the remaining vector pairs is associated to a ``center pair'' such that after certain operations
(such as subtraction) on the vectors, we obtain a pair with a vector difference yielding a lattice vector with a norm less than $R'$.
If we start an iteration with say $N'$ vector pairs and identify $|\centre|$ number of center pairs, then the output consists of $N'-|\centre|$ vector pairs. {An illustration is given in Figure \ref{fig:aks}.} In \cite{2001_AKS} and most other provable variants or generalizations such as \cite{2008_AJ,2009_BN}, the running time of this sieving procedure, 
which is the dominant part of the total running time of the algorithm, is roughly quadratic in the number of sampled vectors.

Here, we propose a different sieving approach to reduce the overall time complexity of the algorithm. This can be thought of as a
generalization of the sieving method introduced in \cite{2018_AM} for the $\ell_{\infty}$ norm. 
We divide the space such that each lattice
vector can be mapped efficiently into some desired division. In the following subsection, we explain this sieving procedure, whose
running time is linear in the number of sampled vectors.

%-----------------------------------------
\subsection{Linear Sieve}
\label{subsec:linSieve_p}

In the initial AKS algorithm \cite{2001_AKS,2002_AKS} as well as in all its variants thereafter \cite{2009_BN,2008_AJ,2010_MV}, in the sieving sub-routine, a space $\ballp_n(R)$ has been divided into sub-regions such that each sub-region is associated with a center. Then, given a vector, we map it to a sub-region and subtract it from the center so that we get a vector of length at most $\gamma R$. We must aim to select these sub-regions such that we can (i) map a vector efficiently to a sub-region (ii) without increasing the number of centers ``too much''. The latter factor is determined by the number of divisions of $\ballp_n(R)$ into these sub-regions and directly contributes to the space (and hence time) complexity. 

\begin{figure}[h]
 \centering
 \includegraphics[width=11cm, height=4cm]{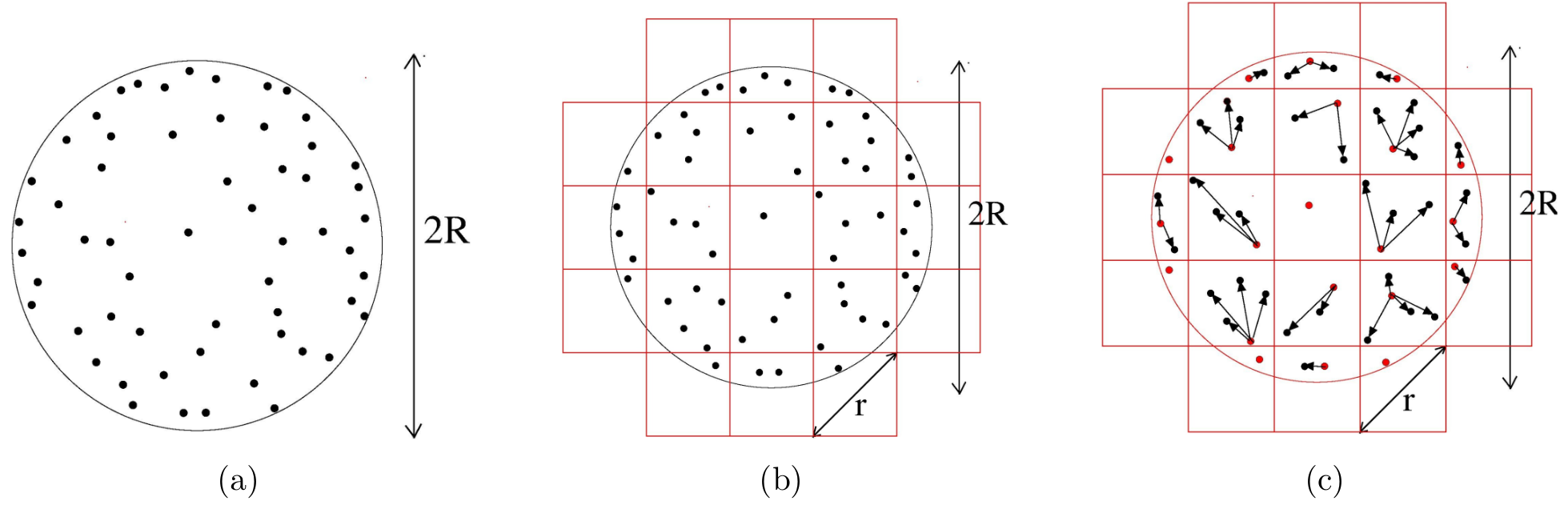}
 \caption{{One iteration of the linear sieve in the $\ell_2$ norm. (\textbf{a}) A number of vector pairs (solid black dots) with (Euclidean) length at most $R$ are sampled. (\textbf{b}) The space is divided into a number of hypercubes with diagonal length $r$, and each vector pair is mapped into a hypercube. (\textbf{c}) Within each hypercube, a subtraction operation (denoted by arrow) is performed between a center (red dot) and the remaining vector pairs, such that we obtain shorter lattice vectors in the next iteration.}}
 \label{fig:linear}
\end{figure}

In all the previous provable sieving algorithms, the sub-regions were small hyperballs (or parts of them) in $\ell_p$ norm. In this paper, our sub-regions are hypercubes. The choice of this particular sub-region makes the mapping very efficient. First, let us note that, in contrast with the previous algorithms (except \cite{2018_AM}), we divide the space a priori. This can be done by dividing each co-ordinate axis into intervals of length $\frac{\gamma R}{n^{1/p}}$ so that the distance between any two vectors in the resulting hypercube is at most $\gamma R$. In an ordered list, we store an appropriate index (say, co-ordinates of one corner) of only those hypercubes which have a non-zero intersection with $\ballp_n(R)$. We can map a vector to a hypercube in $O(n)$ time simply by looking at the intervals in which each of its co-ordinates belong. If the hypercube contains a center, then we subtract the vectors and store the difference; otherwise, we assign this vector as the center. {An illustration is given in Figure \ref{fig:linear}.}

The following lemma gives a bound on the number of hypercubes or centers we obtain by this process. Such a volumetric argument can be found in \cite{1999_P}.
\begin{Lemma}
Let $\gamma\in(0,1), R\in\real_{\geq 1}, 1\leq p\leq\infty$ and $r=\frac{\gamma R}{2n^{1/p}}$. The number of translates of $\balli_n(r)$ required to cover $\ballp_n(R)$ is at most $O\left(\left(2+\frac{2}{\gamma}\right)^n\right)$.

 \label{lem:hypercubeCentre}
\end{Lemma}

\begin{proof}
 Let $N_h$ be the number of translates of $L=\balli_n(r)$ required to cover $K=\ballp_n(R)$. These translates are all within $K\oplus 2L$. In addition, noting that $L \subseteq \frac{rn^{1/p}}{R} K$, we have
 \begin{eqnarray}
  N_h*\vol(L) \leq \vol(K+2L)\leq \left(1+\frac{2rn^{1/p}}{R} \right)^n \vol(K) \nonumber
 \end{eqnarray}
Plugging in the value of $r$, we have $N_h \leq (1+\gamma)^n\frac{\vol(K)}{\vol(L)}$.

Using Fact $\ref{fact:ballVol}$, we have $N_h \in O\left(\left(2+\frac{2}{\gamma}\right)^n\right)$.
\end{proof}

Note that the above lemma implies a sub-division where one hypercube is centered at the origin. Thus, along each axis, we can have the following $2r$-length intervals: $$\ldots [-5r,-3r),[-3r,-r),[-r,r),[r,3r),[3r,5r),\ldots $$
We do not know whether this is the most optimal way of sub-dividing $\ballp_n(R)$ into smaller hypercubes. In \cite{2018_AM}, it has been shown that if we divide $[-R,R]$ from one corner---i.e., place one small hypercube at one corner of the larger hypercube $\balli_n(R)$---then $O\left(\left(\Big\lceil\frac{2}{\gamma}\Big\rceil\right)^n\right)$ copies of hypercubes of radius $r$ suffices. 

Suppose in one sieving iteration, we have a set $S$ of lattice vectors of length at most $R$; i.e., they all lie in $\ballp_n(R)$ {(Figure 3(a))}. We would like to combine points so that we are left with vectors in $\ballp_n(\gamma R)$. We divide each axis into intervals of length $y=\frac{\gamma R}{n^{1/p}}$ and store in an ordered set ($\mathcal{I}$) co-ordinates of one corner of the resulting hypercubes that have a non-zero intersection with $\ballp_n(R)$ {(Figure 3(b))}. {Note} that this can be done in a time of $O(nN_h)$, where $N_h$ is the maximum number of hypercube translates as described in Lemma \ref{lem:hypercubeCentre}. 

We maintain a list $\centrep$ of pairs, where the first entry of each pair is an $n$-tuple in $\mathcal{I}$ (let us call it ``index-tuple'') and the second one, initialized as empty set, is for storing a center pair. Given $\vect{y}$, we map it to its index-tuple $I_{\vect{y}}$ as follows: we calculate the interval in which each of its co-ordinates belong {(steps 10-13 in Algorithm 2)}. This can be done in $O(n)$ time. {This is equivalent to storing information about the hypercube (in Figure 3(b)) in which it belongs or is mapped to.} 
We can access $\centrep[I_{\vect{y}}]$ in constant time. For each $(\vect{e},\vect{y})\in S$, if there exists a $(\vect{e}_{\vect{c}},\vect{c}) \in \centrep[I_{\vect{y}}]$---i.e., 
$I_{\vect{y}}=I_{\vect{c}}$ (implying $\| \vect{y}-\vect{c} \|_p \leq \gamma R $)---then we add 
$(\vect{e},\vect{y}-\vect{c}+\vect{e}_{\vect{c}})$ to the output set $S'$ {(Figure 3(c))}.
Otherwise, we add vector pair $(\vect{e},\vect{y})$ to $\centrep[I_{\vect{y}}]$ as a center pair. {This implies that if there exists a center in the hypercube, then we perform subtraction operations to obtain a shorter vector. Otherwise, we make $(\vect{e},\vect{y})$ the center for its hypercube.}
Finally, we \mbox{return $S'$.}

More details of this sieving procedure (Linear Sieve) can be found in Algorithm \ref{alg:multSieve}.

%-------------------------------------------------------------
\subsection{AKS Algorithm with a Linear Sieve}
\label{subsec:aks_lin}

Algorithm \ref{alg:multI} describes an exact algorithm for $\svpp$ with a linear sieving procedure (Linear Sieve) (Algorithm \ref{alg:multSieve}).

%-------------------------------------------------------------
\begin{algorithm}
 \caption{An exact algorithm for $\svpp$}
 \setlength{\algoheightrule}{1pt} % thickness of the rules above and below
\setlength{\algotitleheightrule}{0.5pt}

 \label{alg:multI}
 
 \KwIn{(i) A basis $\vect{B} = [\vect{b}_1, \ldots \vect{b}_n]$ of a lattice $\cL$, 
 (ii) $ 0 < \gamma <1 $, (iii) $\xi > 1/2$, 
 (iv) $\lambda \approx \minp_1(\cL)$ ,(v) $N \in \nat$ }
 \KwOut{A shortest vector of $\cL$ }
 
 $S \leftarrow \emptyset$ \;
 \For{$i=1$ to $N$  \label{sample:start}}
 {
    $\vect{e}_i \leftarrow_{\text{uniform}} \ballp_n(\vect{0},\xi\lambda) $ \; \label{sample:v}
    $ \vect{y}_i \leftarrow \vect{e}_i \mod \fpar(\vect{B}) $ \;  \label{sample:y}
   % $(\vect{e}_i,\vect{y}_i) \leftarrow \text{Sample}(\vect{B},\xi\lambda)$ 
   % using Algorithm \ref{alg:sample}   \label{multI:sample} \; 
    $S \leftarrow S \cup \{ (\vect{e}_i,\vect{y}_i) \}$	\; 
 }	\label{sample:end}
 $R \leftarrow n \max_i \|\vect{b}_i\|_p $ \;
 
 \For{$j = 1$ to $k= \Big\lceil \log_{\gamma} \Big( \frac{\xi}{nR(1-\gamma)} \Big) \Big\rceil$ }
 {
    $S \leftarrow \text{sieve}(S,\gamma,R,\xi)$ using Linear Sieve (Algorithm \ref{alg:multSieve}) \; \label{multI:sieve}
    $R \leftarrow \gamma R + \xi \lambda$ 	\label{multI:R}	\; 
 }	\label{multI:sieveEnd}
 
    Compute the non-zero vector $\vect{v}_0$  in  
    $ \{ (\vect{y}_i-\vect{e}_i)-(\vect{y}_j-\vect{e}_j) : (\vect{e}_i,\vect{y}_i),(\vect{e}_j,\vect{y}_j) 
    \in S \}$  with the smallest $\ell_p$ norm \label{multI:short} \; 
    \Return $\vect{v}_0$ \;

\end{algorithm}

%---------------------------
\begin{Lemma}
  Let $\gamma \in \real_{(0,1)}$. The number of center pairs in Algorithm \ref{alg:multSieve} always 
  satisfies $|\centrep| \leq 2^{\ccp n+o(n)}$ where $\ccp = \log \left(2+ \frac{2}{\gamma} \right)$.
  
 \label{lem:multCentre}
\end{Lemma}

\begin{proof}
 This follows from Lemma \ref{lem:hypercubeCentre} in Section \ref{subsec:linSieve_p}.
 
\end{proof}

\begin{claim}
 The following two invariants are maintained in Algorithm \ref{alg:multI}: \\
% \begin{enumerate}
  1. $\quad \forall (\vect{e},\vect{y}) \in S, \quad \vect{y} - \vect{e} \in \cL \qquad \qquad $
  2. $\quad \forall (\vect{e}, \vect{y}) \in S, \quad \|\vect{y}\|_p \leq R$.
 %\end{enumerate}
\label{claim:app_multI_invariant}
\end{claim}

\begin{proof}
\begin{enumerate}
 \item The first invariant is maintained at the beginning of the sieving iterations in Algorithm \ref{alg:multI} 
 due to the choice of $\vect{y}$ at step \ref{sample:y} of Algorithm \ref{alg:multI}.
 
 Since each center pair $(\vect{e}_{\vect{c}},\vect{c}) $ once belonged to $S$, $\vect{c} - \vect{e}_{\vect{c}} \in \cL$.
 Thus, at step \ref{multSieve:reduce} of the sieving procedure (Algorithm \ref{alg:multSieve}), we have 
 $(\vect{e}-\vect{y})+(\vect{c}-\vect{e}_{\vect{c}}) \in \cL$. 
 
 \item The second invariant is maintained in steps \ref{sample:start}--\ref{sample:end} of Algorithm \ref{alg:multI} because 
 $\vect{y} \in \fpar(\vect{B})$ and hence 
 $\|\vect{y}\|_p \leq \sum_{i=1}^n \|\vect{b}_i\|_p \leq n \max_i \|\vect{b}_i\|_p = R$.
 
 We claim that this invariant is also maintained in each iteration of the \linebreak sieving procedure.
 
 Consider a pair $(\vect{e},\vect{y}) \in S$ and let $I_{\vect{y}}$ be its index-tuple.
 Let $(\vect{e}_{\vect{c}},\vect{c})$ be its associated center pair.  
 By Algorithm \ref{alg:multSieve}, we have $I_{\vect{y}} = I_{\vect{c}}$; i.e., 
 $\|\vect{y}-\vect{c}\|_p^p = \sum_{i=1}^n |y_i-c_i|^p \leq \sum_{i=1}^n \frac{\gamma^p R^p}{n} \leq \gamma^pR^p$.
 Thus, $\|\vect{y}-\vect{c}\|_p \leq \gamma R$ and hence
 $\quad
  \|\vect{y} - \vect{c} + \vect{e}_{\vect{c}} \|_p \leq 
  \|\vect{y} - \vect{c} \|_p + \|\vect{e}_{\vect{c}} \|_p
  \leq \gamma R + \xi \lambda . $
  
 The claim follows by the re-assignment of variable $R$ at step \ref{multI:R} in Algorithm \ref{alg:multI}.

\end{enumerate}
 
\end{proof}

%------------------------------------------

\begin{algorithm}
 \caption{Linear Sieve for $\ell_p$ norm}
 \label{alg:multSieve} 
 \KwIn{(i) Set $S = \{ (\vect{e}_i,\vect{y}_i) : i \in I \} \subseteq \ballp_n(\xi\lambda) \times \ballp_n(R)$ such that
 $\forall i \in I, \quad \vect{y}_i-\vect{e}_i \in \cL$,
 (ii) $(\gamma,R,\xi)$ }
 \KwOut{A set $S' = \{ (\vect{e'}_i,\vect{y'}_i) : i \in I' \} \subseteq \ballp_n(\xi\lambda) \times 
 \ballp_n(\gamma R+\xi\lambda)$ such that  $ \forall i \in I', \quad \vect{y'}_i-\vect{e'}_i \in \cL$ }
 
 $R \leftarrow \max_{(\vect{e,y}) \in S} \|\vect{y}\|_{p} $ \;
% $\ell = \Big\lceil \frac{2}{\gamma} \Big\rceil$, $\quad \ell_2=\Big\lceil\frac{1-1/n^{1/p}}{\gamma}\Big\rceil$	\;
 $ S' \leftarrow \emptyset$ \;
 Divide each axis into intervals of length $\frac{\gamma R}{n^{1/p}}$ and store a corner of those resulting hypercubes
 with a non zero intersection with $\ballp_n(R)$ in ordered set $\mathcal{I}$ \;    \label{multSieve:division}
 $\centrep\leftarrow\{((i_1,i_2,\ldots,i_n),\emptyset):(i_1,i_2,\ldots,i_n) \in\mathcal{I} \}$ \;
 
 \For{$(\vect{e},\vect{y}) \in S$}
 {
    \eIf{$\|\vect{y}\|_p \leq \gamma R$}
    {
      $S' \leftarrow S' \cup \{ (\vect{e}, \vect{y}) \}$ \;
    }
    {
      $I \leftarrow \emptyset$	\;
      \For{$i=1,\ldots,n$}
     {
        Find the integer $j$ such that $(j-1) \leq \frac{y_i+R}{\gamma R/n^{1/p}} < j$	\;
        $I[i] = j$	\;
      }
      
      \eIf{$\exists (\vect{e}_{\vect{c}},\vect{c}) \in \centrep[I]$ 	 \label{multSieve:compare}}
      {
        $S' \leftarrow S' \bigcup \{(\vect{e},\vect{y}-\vect{c}+\vect{e}_{\vect{c}}) \}$  \label{multSieve:reduce}	\;
      }
	  {
        $\centrep[I] \leftarrow \centrep[I] \bigcup \{ (\vect{e},\vect{y}) \}$	\;	\label{multSieve:addCentre}
      }
    }
 }
 \Return $S'$	\;
\end{algorithm}

%------------------------------------
%--------------------------------------------------

In the following lemma, we bound the length of the remaining lattice vectors after all the sieving iterations \mbox{are over.}
The proof is similar to that given in \cite{2018_AM2}, so we write \mbox{it briefly.}
%------------Final Radius-------------
\begin{Lemma}
 At the end of $k$ iterations in Algorithm \ref{alg:multI}, the length of lattice vectors 
 $ \|\vect{y}-\vect{e}\|_p \leq \frac{\xi (2-\gamma)\lambda}{1-\gamma}+\frac{\gamma \xi}{n(1-\gamma)} =: R'$.
 \label{lem:multI_finRad}
\end{Lemma}

\begin{proof}
 Let $R_k$ be the value of $R$ after $k$ iterations, where \\
 $\log_{\gamma} \Big( \frac{\xi}{nR(1-\gamma)} \Big) \leq k \leq \log_{\gamma} \Big( \frac{\xi}{nR(1-\gamma)} \Big) + 1$.
 
 Then,
 \begin{eqnarray}
  R_k &=& \gamma^k R + \sum_{i=1}^k \gamma^{k-1} \xi \lambda  
    \leq \frac{\xi \gamma}{n(1-\gamma)} + \frac{\xi \lambda}{1-\gamma} \Big[ 1-\frac{\xi}{nR(1-\gamma)} \Big] 
    \nonumber
 \end{eqnarray}
Thus, after $k$ iterations, $\| \vect{y}\|_p \leq R_k$, and hence after $k$ iterations,
\begin{eqnarray}
 \| \vect{y}-\vect{e}\|_p &\leq& \| \vect{y}\|_p + (\|-\vect{e}\|_p)  
    \leq R_k + \xi \lambda \nonumber \\
     &=& \frac{(2-\gamma)\xi \lambda}{1-\gamma} + \frac{\gamma \xi}{n(1-\gamma)} 		\nonumber
\end{eqnarray}
\end{proof}

Using Corollary \ref{cor:latBall} and assuming $\lambda \approx \minp_1$, we obtain an upper bound on the number of 
lattice vectors of a length of at most $R'$; i.e., \\ $|\ballp_n(R') \cap \cL| \leq 2^{\cbp n+o(n)} $, where 
$\cbp=\log \left(1 + \frac{2\xi(2-\gamma)}{1-\gamma} \right)$.

%-------------------------------
The above lemma along with the invariants implies that at the beginning of step \ref{multI:short} in Algorithm \ref{alg:multI}, 
we have ``short'' lattice vectors; i.e., vectors with a norm bounded by $R'$.
We want to start with a ``sufficient number'' of vector pairs so that we do not end up with all zero vectors at the end of the 
sieving iterations.
For this, we work with the following conceptual modification proposed by Regev\cite{2009_R1}.
%{{(\hl{O}.Regev. Lecture notes on lattices in computer science. 2009.)}}. Reply : Not required.

Let $\vect{u} \in \cL$ such that $ \|\vect{u}\|_p = \minp_1(\cL) \approx \lambda$ (where $2 < \minp_1(\cL) \leq 3$), 
$D_1 = \ballp_n(\xi \lambda) \cap \ballp_n(-\vect{u},\xi \lambda)$ and
$D_2 = \ballp_n(\xi \lambda) \cap \ballp_n(\vect{u},\xi \lambda)$.
Define a bijection $\sigma$ on $\ballp_n(\xi \lambda)$ that maps $D_1$ to $D_2$, $D_2$ to $D_1$ and $\ballp_n(\xi \lambda) \setminus (D_1 \cup D_2)$ to itself :
\begin{eqnarray}
 \sigma(\vect{e}) = 
      \begin{cases}
       \vect{e} + \vect{u} & \text{ if } \vect{e} \in D_1	\\
       \vect{e} - \vect{u} & \text{ if } \vect{e} \in D_2 	\\
       \vect{e} & \text{ else }
      \end{cases}
\nonumber
\end{eqnarray}
For the analysis of the algorithm, we assume that for each perturbation vector $\vect{e}$ chosen by our algorithm, we replace 
$\vect{e}$ by $\sigma(\vect{e})$ with probability $1/2$ and that it remains unchanged with probability $1/2$. 
We call this procedure {\em tossing} the vector $\vect{e}$. 
This does not change the distribution of the perturbation vectors $\{\vect{e}\}$.
Further, we assume that this replacement of the perturbation vectors happens at the step where this has any effect on the algorithm  for the first time. 
In particular, at step \ref{multSieve:addCentre} in Algorithm \ref{alg:multSieve}, after we have identified a center pair 
$(\vect{e}_c,\vect{c})$, we apply $\sigma$ on $\vect{e}_c$ with probability $1/2$.
Then, at the beginning of step \ref{multI:short} in Algorithm \ref{alg:multI}, we apply $\sigma$ to $\vect{e}$ 
for all pairs $(\vect{e}, \vect{y}) \in S$.
The distribution of $\vect{y}$ remains unchanged by this procedure because $\vect{y} \equiv \vect{e} \equiv \sigma(\vect{e})\mod 
\fpar(\vect{B})$ and $\vect{y}-\vect{e} \in \cL$.
A somewhat more detailed explanation of this can be found in the following result of \cite{2009_BN}.
\begin{Lemma}[\textbf{Theorem 4.5 in } \cite{2009_BN} (re-stated)]
 The modification outlined above does not change the output distribution of the actual procedure.
\end{Lemma}
Note that since this is just a conceptual modification intended for ease in analysis, we should not be concerned with the 
actual running time of this modified procedure.
Even the fact that we need a shortest vector to begin the mapping $\sigma$ does not matter.

The following lemma will help us to estimate the number of vector pairs to sample at the beginning of the 
algorithm.
\begin{Lemma}[\textbf{Lemma 4.7 in } \cite{2009_BN}]
 Let $N \in \nat$ and $q$ denote the probability that a random point in $\ballp_n(\xi \lambda)$ is contained in 
 $D_1 \cup D_2$.
 If $N$ points $\vect{x}_1, \ldots \vect{x}_N$ are chosen uniformly at random in $\ballp_n(\xi \lambda)$, 
 then with a probability larger than $1-\frac{4}{qN}$, there are at least $\frac{qN}{2}$ points 
 $\vect{x}_i \in \{ \vect{x}_1, \ldots \vect{x}_N \}$ with the property $\vect{x}_i \in D_1 \cup D_2$.
 \label{lem:multI_goodPair}
\end{Lemma}
From Lemma \ref{lem:overlap}, we have 
\begin{eqnarray}
q &\geq& 2^{-\csp n}  \qquad  \text{where } \csp = -\log\Big( 0.5 - \frac{1}{4\xi} \Big)
\nonumber
\end{eqnarray}
Thus, with a probability of at least $1-\frac{4}{qN}$, we have at least $ 2^{-\csp n} N $ pairs $(\vect{e}_i,\vect{y}_i)$ before the 
sieving iterations such that $ \vect{e}_i \in D_1 \cup D_2$.

%----------------Time complexity----------------
\begin{Lemma}
  If $N \geq \frac{2}{q}(k|\centrep| + 2^{\cbp n} +1)$, then with a probability of at least $1/2$, Algorithm \ref{alg:multI} outputs 
  a shortest non-zero vector in $\cL$ with respect to $\ell_p$ norm for $1\leq p \leq \infty$.
 \label{lem:multI_zero}
\end{Lemma}

\begin{proof}
 Of the $N$ vector pairs $(\vect{e},\vect{y})$ sampled in steps \ref{sample:start}--\ref{sample:end} of Algorithm \ref{alg:multI}, 
 we consider those such that $\vect{e} \in (D_1 \cup D_2)$. 
 We have already seen there are at least $\frac{qN}{2}$ such pairs with a probability of at least $1-\frac{4}{qN}$.
 We remove $|\centrep|$ vector pairs in each of the $k$ sieve iterations.
 Thus, at step \ref{multI:short} of Algorithm \ref{alg:multI}, we have $N' \geq 2^{\cbp n}+1$ pairs $(\vect{e},\vect{y})$ 
 to process.
 
 By Lemma \ref{lem:multI_finRad}, each of them is contained within a ball of radius $R'$ which can have at most $2^{\cbp n}$ lattice
 vectors.
 Thus, there exists at least one lattice vector $\vect{w}$ for which the perturbation is in $D_1 \cup D_2$, and 
 it appears twice in $S$ at the beginning of step \ref{multI:short}. 
 With a probability of $1/2$, it remains $\vect{w}$, or with the same probability, it becomes either $\vect{w}+\vect{u}$ or
 $\vect{w}-\vect{u}$.
 Thus, after taking pair-wise difference at step \ref{multI:short} with a probability of at least $1/2$, we find the shortest vector.
\end{proof}

\begin{Theorem}
  Let $\gamma \in (0,1)$, and let $\xi >1/2$. 
  Given a full rank lattice $\cL \subset \ratn^n$, there is a randomized algorithm for $\svpp$ with a success probability of at 
  least $1/2$, a space complexity of at most $2^{\cspacep n+o(n)}$, and running time of at most $2^{\ctimep n+o(n)}$,
  where $\cspacep = \csp + \max(\ccp,\cbp)$ and $\ctimep = \max(\cspacep,2\cbp)$, where \\
  $\ccp =\log \left(2+\frac{2}{\gamma} \right), \quad 
  \csp = -\log\Big( 0.5 - \frac{1}{4\xi} \Big) $ and
  $\cbp=\log \left(1 + \frac{2\xi(2-\gamma)}{1-\gamma} \right)$.
  
 % In particular for $\gamma=0.41$ and $\xi=0.742$ the algorithm runs in time $2^{4n+o(n)}$ with a corresponding space 
 % requirement of at most $2^{3.938n+o(n)}$.
  
 \label{thm:multI}
\end{Theorem}

\begin{proof}
 If we start with $N$ pairs (as stated in Lemma \ref{lem:multI_zero}), then the space complexity is at most 
  $2^{\cspacep n + o(n)}$ with $\cspacep = \csp + \max (\ccp,\cbp)$.
 
 In each iteration of the sieving Algorithm \ref{alg:multSieve}, it takes at most $O(nN_h)$ time to initialize and index 
 $\centrep$ (Lemmas \ref{lem:hypercubeCentre} and \ref{lem:multCentre}).
 For each vector pair $(\vect{e},\vect{y}) \in S$, it takes a time of at most $n$ to calculate its index-tuple $I_{\vect{y}}$.
 Thus, the time taken to process each vector pair is at most $(n+1)$, and the total time taken per iteration of Algorithm \ref{alg:multSieve} is at most 
 $O(n(N_h+N))$,  which is at most $2^{\cspacep n + o(n)}$, and there are at most $\poly(n)$ such iterations.
  
 If $N' \geq 2^{\cbp n}+1$, then the time complexity for the computation of the pairwise differences is at most 
 $(N')^2 \in 2^{2\cbp n+o(n)}$.
 
 Thus, the overall time complexity is at most $2^{\ctimep n + o(n)}$ where \\
 $\ctimep = \max (\cspacep,2\cbp)$.
\end{proof}

%---------------------------------------------------------------------
%	BIRTHDAY PARADOX
%----------------------------------------------------------------------
\subsection{Improvement Using the Birthday Paradox}
\label{svpi:bday}

We can obtain a better running time and space complexity if we use the birthday paradox to decrease the number of sampled vectors but obtain at least two vector pairs corresponding to the same lattice vector after the sieving iterations \cite{2009_PS,2011_HPS}. For this, we have to ensure that the vectors are independent and identically distributed before step \ref{multI:short} of Algorithm \ref{alg:multI}. Thus, we incorporate the following modification{, as discussed in \cite{2011_HPS}. Very briefly, the trick is to set aside many uniformly distributed vector pairs as centers for each sieving step, even before the sieving iterations begin. In each sieving iteration, the probability that a vector pair is not within the required distance of any center pair decreases. Now, if we sample enough vectors, then with a good probability at step \ref{multI:short}, we have at least two vectors whose perturbation is in $D_1\bigcup D_2$, implying that with a probability of at least 1/2, we obtain the \mbox{shortest vector.}}

{In the analysis of \cite{2011_HPS}, the authors simply stated that the required center pairs can be sampled uniformly at the beginning. In our linear sieving algorithm, we have an advantage. Unlike the AKS-style algorithms, in which the center pairs are selected and then the space is divided, in our case, we can divide the space a priori. We take advantage of this and conduct a number of random divisions of the space. Since in each iteration, the length of the vectors decreases, the size of the hypercubes also decreases, and this can be calculated. Thus, for each iteration we have a number of divisions of the space into hypercubes of a certain size. For this, we need to divide the axes into intervals of a fixed size. Simply by shifting the intervals in each axis, we can make this division random. Then, among the uniformly sampled vectors, we select a center for each hypercube.}

Assume we start with $N \geq \frac{2}{q}(n^3 k|\centrep| + n 2^{\frac{\cbp}{2}n})$ sampled pairs. 
After the initial sampling, for each of the $k$ sieving iterations, we fix $\Omega\Big(\frac{2n^3}{q} |\centrep|\Big)$ pairs to be used as center pairs in the following way.

{1.} Let $R = \max_{i \in [N]} \|\vect{y}_i\|_p $. We maintain $k$ lists of pairs, $\centrep_1,\centrep_2,\ldots,\centrep_k$, where each list is similar to ($\centrep$), as described in Algorithm \ref{alg:multSieve}. In the $i^{th}$ list, we store the indices (co-ordinates of a corner) of translates of $\balli_n(r_i)$ that have a non-zero intersection with $\ballp_n(R_i)$ where $R_i=\gamma^{i-1}R+\xi\lambda\frac{1-\gamma^{i-1}}{1-\gamma}$ and $r_i=\frac{\gamma R_i}{2n^{1/p}}$.
For such a division, we can obtain $O(|\centrep|)$ center pairs in each list. 
To meet our requirement, we maintain $O(n^3)$ such lists for each $i$. 
%Note we can have different $2 r_i$ length divisions of a certain interval, depending on the starting point. So in each such group for each list we chose a random starting point and divide the respective intervals as before. 
We call these $O(n^3)$ lists the ``sibling lists'' of $\centrep_i$.

{2. }For each $(\vect{e},\vect{y}) \in S$ {(where $S$ is the set of sampled pairs)}, we first calculate $\|\vect{y}\|_p$ to check in which list group it can potentially 
belong, say $\centrep_j$. That is, $\centrep_j$ corresponds to the smallest hyperball containing $\vect{y}$.
Then, we map it to its index-tuple $I_{\vect{y}}$, as has already been described before.
We add $(\vect{e},\vect{y})$ to a list in $\centrep_j$ or any of its sibling lists if it was empty before.
%else we subtract vectors as in step \ref{multSieve:reduce} of Algorithm \ref{alg:multSieve}.
Since we sampled uniformly, this ensures we obtain the required number of (initially) fixed centers, and no other vector can be used as a center throughout the algorithm.

{Having set aside the centers,} now we repeat the following sieving operations $k$ times. For each vector pair $(\vect{e}_1,\vect{y}_1) \in S$, we can check which list (or its sibling lists) it can belong to from $\|\vect{y}_1\|_p$. Then, if a center pair is found, we subtract as in step \ref{multSieve:reduce} of Algorithm \ref{alg:multSieve}. Otherwise, we discard it and consider it ``lost''.

Let us call this modified sieving procedure \textbf{{LinearSieveBirthday}%Is the bold necessary? Reply : Yes.
}.
 We obtain the following improvement in the running time.
\begin{Theorem}
Let $\gamma \in (0,1)$, and let $\xi >1/2$. 
Given a full rank lattice $\cL \subset \ratn^n$, there is a randomized algorithm for $\svpp$ with a success probability of at 
least $1/2$, a space complexity of at most $2^{\cspacep n+o(n)}$, and running time of at most $2^{\ctimep n+o(n)}$,
where $\cspacep = \csp + \max(\ccp,\frac{\cbp}{2})$ and $\ctimep = \max(\cspacep,\cbp)$, where \\
$\ccp =\log \left(2+\frac{2}{\gamma} \right), \quad \csp = - \log \left(0.5-\frac{1}{4\xi} \right)$ and
$\cbp = \log \left(1 + \frac{2\xi(2-\gamma)}{1-\gamma} \right)$.
  
 \label{thm:multI_bday}
\end{Theorem}

{\begin{proof} This analysis has been taken from \cite{2011_HPS}. At the beginning of the algorithm, among the pairs set aside as centers for the first step, there are $\Omega\left(n^3|\centre|\right)$ pairs such that the perturbation is in $D_1\bigcup D_2$ with high probability (Lemma \ref{lem:multI_goodPair}). We call them good pairs. After fixing these pairs as centers, the probability that the distance between the next perturbed vector and the closest center is more than $\gamma R$ decreases. The sum of these probabilities is bounded from above by $|\centre|$. As a consequence, once all centers have been processed, the probability for any of the subsequent pairs to be lost is $O\left(\frac{1}{n^3}\right)$. By induction, it can be proved that the same proportion of pairs is lost at each step of the sieve with high probability. As a consequence, no more than $1-\left(1-\frac{1}{n^3}\right)^{O(n^2)}=O\left(\frac{1}{n}\right)$ pairs are lost during the whole algorithm. This means that in the final ball, there are $\Omega\left(n2^{\frac{c_b}{2}n}\right)$ probabilistically independent lattice points corresponding to good pairs with high probability. As in the proof of Lemma \ref{lem:multI_zero} this implies that the algorithm returns a shortest vector with a probability of at least 1/2.
\end{proof} }

\subsection*{{Comparison} %Please confirm whether this is a secondary title or a list, if it is a secondary title, please modify it. And please confirm if it’s a list, italics is necessary  Reply : It is a list and italics are necessary.
of Linear Sieve with provable sieving algorithms \cite{2001_AKS,2002_AKS,2009_BN,2008_AJ}}
\label{subsec:compare}

For $1\leq p \leq\infty$, the number of centers obtained by \cite{2009_BN} is \\$|\centrep(BN)|\leq 2^{\ccp(BN)n}$, where
$\ccp(BN)=\log(1+\frac{2}{\gamma})$ (Lemma \ref{lem:cc2}). If we conducted a similar analysis for their algorithm, we would obtain space and time complexities of $2^{\cspacep(BN)n+o(n)}$ and $2^{\ctimep(BN)n+o(n)}$, respectively, where
\begin{eqnarray}
\cspacep(BN)&=&\csp+\max(\ccp(BN),\cbp) \nonumber \\
\text{ and }\ctimep(BN)&=&\max(\cspacep(BN)+\ccp(BN),2\cbp).    \nonumber
\end{eqnarray} 
{We} %mdpi:Please confirm whether indentation needs to be added here Reply : No
can incorporate modifications to apply the birthday paradox, as has been done in \cite{2011_HPS} (for $\ell_2$ norm). This would improve the exponents to 
\begin{eqnarray}
\cspacep(BN')&=&\csp+\max(\ccp(BN),\cbp/2) \nonumber\\
\text{ and }\ctimep(BN')&=&\max(\cspacep(BN)+\ccp(BN),\cbp).    \nonumber
\end{eqnarray}
{Clearly}, the running time of our algorithm is less since $\left(1+\frac{2}{\gamma}\right)^2 > \left(2+\frac{2}{\gamma}\right)$ for all $\gamma<1$.
In \cite{2009_BN}, the authors did not specify the constant in the exponent of running time. However, using the above formulae, we found out that their algorithm can achieve a time complexity of $2^{3.849n+o(n)}$ and space complexity of $2^{2.023n+o(n)}$ at parameters $\gamma=0.78,\xi=1.27$ {({without} the birthday paradox, the algorithm in \cite{2009_BN} can achieve time and space complexities of $2^{5.179n+o(n)}$ and $2^{3.01n+o(n)}$, respectively, at parameters $\gamma=0.572,\xi=0.742$)}. In comparison, our algorithm can achieve a time and space complexity of $2^{2.751n+o(n)}$ at parameters \mbox{$\gamma=0.598,\xi=0.82$.}

For $p=2$, we can use Theorem \ref{thm:KL} to obtain a better bound on the number of lattice vectors that remain after all sieving iterations. This is reflected in the quantity $\cbp$, which is then given by $\cb=0.401+\log\Big(\frac{2\xi(2-\gamma)}{1-\gamma}\Big)$ (Corollary \ref{cor:latBall}). Furthermore, $\cs=-0.5\log\Big(1-\frac{1}{4\xi^2}\Big)$ (Lemma \ref{lem:overlap}). 
At parameters $\gamma=0.693$ and $\xi=0.99$, we obtain $\ctime=\cspace=2.49$. The AKS algorithm with the birthday paradox manages to achieve a time complexity of $2^{2.571n+o(n)}$ and space complexity of $2^{1.407n+o(n)}$ when $\gamma=0.589$ and $\xi=0.9365$ \cite{2011_HPS}. 
Thus, our algorithm achieves a better time complexity at the cost of more space. 

For $p=\infty$, we can reduce the space complexity by using the sub-division mentioned in Section \ref{subsec:linSieve_p} and achieve a space and time complexity of $2^{2.443n+o(n)}$ at parameters \mbox{$\gamma=0.501,\xi=0.738$} {(in \cite{2018_AM}, the authors {mentioned} a time and space complexity of $2^{2.82n+o(n)}$ in $\ell_{\infty}$ norm. We obtain a slightly better running time by using $\cbp$, as mentioned in this paper)}.
Again, this is better than the time complexity of \cite{2009_BN} (which is for all $\ell_p$ norms).
\section{A Mixed Sieving Algorithm}
\label{sec:subQuad}

The main advantage in dividing the space (hyperball) into hypercubes (as we did in Linear Sieve) is the efficient ``decodability'' in the sense that a vector can be mapped to a sub-region (and thus be associated with a center) in $O(n)$ time. However, the price we pay is in space complexity, because the number of hypercubes required to cover a hyperball is greater than the number of centers required if we used smaller hyperballs like in~\cite{2001_AKS,2009_BN,2008_AJ}. To reduce the space complexity, we perform a mixed sieving procedure. Double sieving techniques have been used for heuristic algorithms as in \cite{2011_WLTB}, where the rough idea is the following. There are two sets of centers: the first set consists of centers of larger radius balls, and for each such center, there is another set of centers of smaller radius balls within the respective large ball. In each sieving iteration, each non-center vector is mapped to the larger balls by comparing with the centers in the first set. Then, they are mapped to a smaller ball by comparing with the second set of centers. Thus, in both levels, a quadratic sieve \mbox{is applied. }

\begin{figure}[h]
 \centering
 \includegraphics[width=11cm, height=4cm]{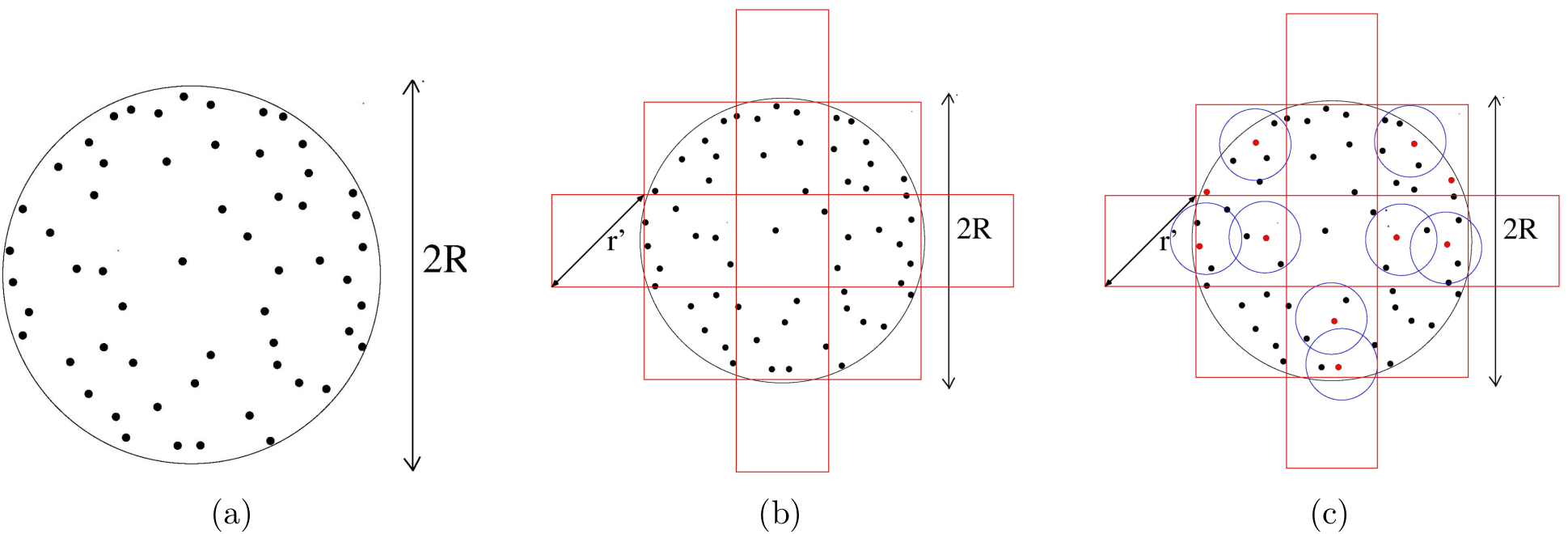}
 \caption{{One iteration of the mixed sieve in the $\ell_2$ norm. (\textbf{a}) A number of vector pairs (solid black dots) with a (Euclidean) length of at most $R$ are sampled. (\textbf{b}) The space is divided into hypercubes with diagonal length $r'$, and the vector pairs are mapped into each hypercube. (\textbf{c}) Within each hypercube, some vector pairs are selected as centers (red dots) and a hypercube is further sub-divided into a number of $\ell_2$ balls, centered around these red dots. Then, vector subtraction is performed between the center and the vector pairs in each $\ell_2$ ball (like AKS).}}
 \label{fig:mix}
\end{figure}

In our mixed sieving, the primary difference is the fact that in the two levels, we use two types of sieving methods: a linear
sieve in the first level and then a quadratic sieve such as AKS in the next level. The overall outline of the algorithm is the same as in \mbox{Algorithm \ref{alg:multI}}, except at step \ref{multI:sieve}, where we apply the following sieving procedure, which we call Mixed Sieve. {An illustration is  given in Figure \ref{fig:mix}.}

The input to Mixed Sieve is a set of vectors of length $R$, and the output is a set of smaller vectors of length $\gamma R$.
\begin{enumerate}
 \item We divide the whole space into large hypercubes of length $\frac{A\gamma R}{n^{1/p}}$, where $A$ is some constant. 
 In $O(n)$ time, we map a vector to a large hypercube by comparing its co-ordinates. This has been explained in Section \ref{subsec:linSieve_p}. We do not assign centers yet and do not perform any vector operation at this step. The distance between any two vectors mapped to the same hypercube is at most $A\gamma R$ {(Figure \ref{fig:mix}b)}.
 
 \item Next, we perform the AKS sieving procedure within each hypercube. For each hypercube, we have a set (initially null) of centers. When a vector is mapped to a hypercube ,we check if it is within distance $\gamma R$ of any center (within that hypercube). If yes, then we subtract it from the center and add the resultant shorter vector to output set. If no, then we add this vector to the set of centers {(Figure \ref{fig:mix}c)}.
 
\end{enumerate}

Using the same kind of counting method as in Section \ref{subsec:linSieve_p}, we can say we need $2^{c' n}$ large hypercubes, where 
$c'=\log \left(2+\frac{2}{A\gamma}\right)$. The maximum distance between any two vectors in each hypercube is 
$A\gamma R$, and we want to get vectors of length at most $\gamma R$ by applying the AKS sieve. Thus, the number of centers 
(let us call these ``AKS sieve-centers'') within each hypercube is $2^{c_p n+o(n)}$ where $c_p=\log(1+A)$ (in the special case of Euclidean norm, we have $c_2=0.401-\log\left(\frac{2}{A}\right)$). $c_p$ (and $c_2$) are obtained by applying Lemma \ref{lem:cc2}. Note that the value of $A$ must ensure the non-negativity of $c_2$. Thus, the total number of centers is $2^{\cmixp n+o(n)}$ where 
$\cmixp=c'+c_p$.

To use the birthday paradox, we apply similar methods as given in Section \ref{svpi:bday} and \cite{2011_HPS}. Assume that we initially
sample $N \geq \frac{2}{q}(n^3 k 2^{\cmixp n+o(n)} + n 2^{\frac{\cbp}{2}n})$ vectors. Then, using similar arguments as in Section \ref{sec:aks_p}, we can conclude that, with high probability, we end up with the shortest vector in the lattice. {We are not re-writing the proof since it is similar to that in \mbox{Theorem \ref{thm:multI_bday}.} The only thing that is slightly different is the number of center pairs set aside at the beginning of the sieving iterations. As in Section \ref{sec:aks_p}, we randomly divide the space $n^3$ times into $2^{c'n}$ hypercubes. Then, among the uniformly sampled vectors, we set aside $2^{c_pn}$ vector pairs as centers for each hypercube. Thus, in Theorem \ref{thm:multI_bday}, we replace $|\centre|$ by $2^{c^{(p)}n+o(n)}$. }

Thus, space complexity is $2^{\cspacep n+o(n)}$ where $\cspacep=\csp+\max(\cmixp,\cbp/2)$. 
%Here $\csp = - 0.5\log \left(1-\frac{1}{4\xi^2} \right)$, as obtained from Lemma \ref{lem:overlap}.
It takes $O(n)$ time to map each vector to a large hypercube, and then at most $2^{c_p n+o(n)}$ time to compare it with the ``AKS sieve-centers'' within each hypercube. Thus, the time complexity is $2^{\ctimep n+o(n)}$ where $\ctimep=\max(\cspacep+c_p,\cbp)$.

\begin{Theorem}
Let $\gamma \in (0,1), \xi >1/2$ and $A$ be some constant.
Given a full-rank lattice $\cL \subset \ratn^n$, there is a randomized algorithm for $\svpp$ with a success probability of at 
least $1/2$, a space complexity of at most $2^{\cspacep n+o(n)}$, and a running time of at most $2^{\ctimep n+o(n)}$.
Here, $\cspacep = \csp + \max(\cmixp,\frac{\cbp}{2})$ and $\ctimep = \max(\cspacep+c_p,\cbp)$.
$\csp=-\log\left(0.5-\frac{1}{4\xi}\right)$, $\cbp = \log \left(1 + \frac{2\xi(2-\gamma)}{1-\gamma} \right)$, $c_p=\log(1+A)$ and 
$\cmixp=\log \left(2+\frac{2}{A\gamma}\right)+c_p$.

In the Euclidean norm, we have $c_2=0.401-\log\left(\frac{2}{A}\right)$, \\$\cmix=\log \left(2+\frac{2}{A\gamma}\right)+c_2$, $\cs = - 0.5\log \left(1-\frac{1}{4\xi^2} \right)$  and $\cb=0.401+\log\Big(\frac{2\xi(2-\gamma)}{1-\gamma}\Big)$.
  
 \label{thm:subQuad}
\end{Theorem}

\subsection*{{Comparison} %Please confirm whether this is a secondary title or a list, if it is a secondary title, please modify it. And please confirm if it’s a list, italics is necessary  Reply : It is a list and italics is necessary
with previous provable sieving algorithms \cite{2010_MV,2009_PS,2015_ADRS}}

In the Euclidean norm with parameters $\gamma=0.645, \xi=0.946$ and $A=2^{0.599}$, we obtain a space and time complexity of $2^{2.25n+o(n)}$, while the  List Sieve Birthday \cite{2011_HPS,2009_PS} has space and time complexities of $2^{1.233n+o(n)}$ and $2^{2.465n+o(n)}$, respectively. We can also use a different sieve in the second level, such as List Sieve \cite{2010_MV}, etc., which works in $\ell_2$ norm and is faster than the AKS sieve. We can therefore expect to achieve a better running time.

The Discrete Gaussian-based sieving algorithm of Aggarwal et al. \cite{2015_ADRS} with a time complexity of $2^{n+o(n)}$ performs better than both our sieving techniques. However, their algorithm works for the Euclidean norm and, to the best of our knowledge, it has not been generalized to any other norm.
%better time complexity than List Sieve Birthday \cite{2009_PS,2011_HPS} (see also Section \ref{subsec:compare}) but space compexity is worse.

%----------------------------------------------------
%\begin{remark}
%This kind of mixed sieving technique can also be applied in $\ell_p$ norm for $1\leq p\leq\infty$, where they can improve the 
%running time as well as space requirement.
%\end{remark}

%\input{./approx}
\section{Approximation Algorithms for $\svpp$ and $\cvpp$}
\label{sec:approx}

In this section, we show how to adopt our sieving techniques to approximation algorithms for $\svpp$ and $\cvpp$. The analysis and explanations are similar to that given in~\cite{2018_AM2}. For completeness, we give a brief outline.

\subsection{Algorithm for Approximate $\svpp$}
\label{sec:svpi-approx}

We note that at the end of the sieving procedure in Algorithm~\ref{alg:multI}, we obtain lattice vectors of length at most $R' = \frac{\xi (2-\gamma)\lambda}{1-\gamma} + O(\lambda/n)$. Thus, if we can ensure that one of the vectors obtained at the end of the sieving procedure is non-zero, we obtain a $\tau = \frac{\xi (2-\gamma)}{1-\gamma} + o(1)$-approximation of the shortest vector. Consider a new algorithm ${\mathcal A}$ (let us call it Approx-SVP) that is identical to Algorithm~\ref{alg:multI}, except that Step~\ref{multI:short} is replaced by the following:
\begin{itemize}
\item Find a non-zero vector $\vect{v}_0$  in $ \{ (\vect{y}_i-\vect{e}_i) : (\vect{e}_i,\vect{y}_i)\in S \}$.
\end{itemize}

We now show that if we start with sufficiently many vectors, we must obtain a non-zero vector.
\begin{Lemma}
  If $N \geq \frac{2}{q}(k|\centrep| +1)$, then with a probability of at least $1/2$, Algorithm $\cA$ outputs a non-zero vector in $\cL$ of a length of at most $\frac{\xi (2-\gamma)\lambda}{1-\gamma} + O(\lambda/n)$ with respect to $\ell_p$ norm.
 \label{lem:multI_zero_approx}
\end{Lemma}

\begin{proof}
 Of the $N$ vector pairs $(\vect{e},\vect{y})$ sampled in steps \ref{sample:start}-\ref{sample:end} of Algorithm $\cA$,  we consider those such that $\vect{e} \in (D_1 \cup D_2)$.  We have already seen there are at least $\frac{qN}{2}$ such pairs.
 We remove $|\centrep|$ vector pairs in each of the $k$ sieve iterations. Thus, at step \ref{multI:short} of Algorithm \ref{alg:multI}, we have $N' \geq 1$ pairs $(\vect{e},\vect{y})$  to process.

 With a probability of $1/2$, $\vect{e}$, and hence $\vect{w} = \vect{y} - \vect{e}$ is replaced by either $\vect{w}+\vect{u}$ or
 $\vect{w}-\vect{u}$. Thus, the probability that this vector is the zero vector is at most $1/2$. 
\end{proof}

We thus obtain the following result.
\begin{Theorem}
 Let $\gamma \in (0,1)$, $\xi >1/2$ and $\tau  = \frac{\xi (2-\gamma)}{1-\gamma} + o(1)$, Assume we are given a full-rank lattice $\cL \subset \ratn^n$. There is a randomized algorithm that $\tau$ approximates $\svpp$ with a success probability of at least $1/2$ and a space and time complexity $2^{(\csp + \ccp) n + o(n)}$, where 
 $\ccp =\log \left(2+\frac{2}{\gamma} \right)$, and $\csp = - \log \left(0.5-\frac{1}{4\xi} \right)$. 
 \label{thm:multI-approx}
\end{Theorem}

%In particular for $\gamma = 2/3 +o(1)$ and $\xi = \tau/4$, the algorithm runs in time  $3^n\cdot\left(\frac{\tau}{\tau-2}\right)^n$. For large enough $\tau$ (say $\Omega(\log n)$ for large enough $n$) and $\gamma\rightarrow 1$ the space requirement and running time of this algorithm is approximately $O(2^{n+o(n)})$.

Note that while presenting the above theorem, we assumed that we are using the Linear Sieve in Algorithm \ref{alg:multI}. We can also use the Mixed Sieve procedure as described in \mbox{Section \ref{sec:subQuad}}. Then, we will obtain space and time complexities of $2^{(\csp+\cmixp) n+o(n)}$ and $2^{(\csp+\cmixp+c_p)n+o(n)}$, respectively, where $\cmixp=\log \left(2+\frac{2}{A\gamma}\right)+c_p$ and $c_p=\log(1+A)$, respectively (in the Euclidean norm, the parameters are as described in Theorem \ref{thm:subQuad}).

\paragraph{{Comparison} %Please confirm whether this is a four-level heading or a list, if it is a four-level heading, please modify it. And please confirm if it’s a list, italics is necessary    Reply : Its is a list and italics is required.
with provable approximation algorithms \cite{2011_LWXZ,2009_BN,2008_AJ}}

We have mentioned in Section \ref{sec:intro} that \cite{2008_AJ,2009_BN} gave approximation algorithms for lattice problems that work for all $\ell_p$ norms and use the quadratic sieving procedure (as has been described before). Using our notations, the space and time complexities of their approximate algorithms are $2^{\cspacep(BN)n+o(n)}$ and $2^{\ctimep(BN)n+o(n)}$, respectively, where
\begin{eqnarray}
\cspacep(BN)&=&\csp+\ccp(BN) \nonumber\\
\text{ and } \ctimep(BN)&=&\cspacep(BN)+\ccp(BN)=\csp+2\ccp(BN).\nonumber
\end{eqnarray}
The authors did not mention any explicit value of the constant in the exponent. Using the above formulae, we conclude that \cite{2008_AJ} and \cite{2009_BN} can achieve time and space complexities of $2^{3.169n+o(n)}$ and $2^{1.586n+o(n)}$, respectively, at parameters $\gamma=0.99,\xi=10.001$ with a large constant approximation factor. In comparison, we can achieve a space and time complexity of $2^{2.001n+o(n)}$ with a large constant approximation factor at the same parameters.

In $\ell_2$ norm, using the mixed sieving procedure, we obtain a time and space complexity of $2^{1.73n+o(n)}$ and a large constant approximation factor at parameters $\gamma=0.999$,$\xi=1$. In \cite{2011_LWXZ}, the best running time reported is $2^{0.802n}$ for a large approximation factor. 

Using a similar linear sieve, a time and space complexity of $3^n$ i.e., $2^{1.585n+o(n)}$ can be achieved for the $\ell_{\infty}$ norm for a large constant approximation factor \cite{2018_AM}.  

\begin{comment}
we obtain a time and space complexity of giving a large constant approximation factor. Using mixed sieving technique we achieve a time and space complexity of $2^{1.207n+o(n)}$ with the same 
parameters. In \cite{2011_LWXZ} the authors have analysed the List Sieve Birthday algorithm and gave better packing bounds which
depends on the ratio $\lambda_2^{(2)}/\lambda_1^{(2)}$. They achieve a time complexity of $2^{0.802n+o(n)}$ when this gap
between the first two minima of the lattice is large enough. Our bounds do not depend on this gap. So with lattices where 
this gap is not large then our algorithm performs better, as evident from the results tabulated in \cite{2011_LWXZ} for different
gap values.
\end{comment}

%----------------------------------------------------
\subsection{Algorithm for Approximate $\cvpp$}
\label{subsec:approx_cvp}

Given a lattice $\cL$ and a target vector $\vect{t}$, let $d$ denote the distance of the closest vector in $\cL$ to $\vect{t}$. 
Just as in Section~\ref{subsec:aks_lin}, we assume that we know the value of $d$ within a factor of $1 + 1/n$. 
We can get rid of this assumption by using Babai's~\cite{1986_B} algorithm to guess the value of $d$ within a factor of $2^n$
and then run our algorithm for polynomially many values of $d$. 

For $\tau > 0$, define the following $(n+1)-$dimensional lattice $\cL'$
 \begin{eqnarray}
  \cL' = \lat\Big( \{ (\vect{v},0):\vect{v} \in \cL \} \cup \{(\vect{t},\tau d/2)\} \Big)\;. 	\nonumber
 \end{eqnarray}
 %Consider the unique $k$ satisfying $(1+\alpha)^k \leq d < (1+\alpha)^{k+1}$
 Let $\vect{z}^* \in \cL$ be the lattice vector closest to $\vect{t}$. \\
 Then $\vect{u} = (\vect{z}^*-\vect{t},-\tau d/2) \in \cL'\setminus (\cL-k'\vect{t},0)$ for some $k' \in \intg$.
 %\begin{eqnarray}
  %\|(\vect{z}^*-\vect{t},-\tau d)\|_{\infty} &=& \max(\|\vect{z}^*-\vect{t}\|_{\infty},|-\tau d|)=\max(d,\tau d) \nonumber
 %\end{eqnarray}
 
 We sample $N$ vector pairs $(\vect{e},\vect{y}) \in \ballp_n(\xi d) \times \fpar(\vect{B}')$ 
 (\ref{sample2:start}--\ref{sample2:end} of Algorithm \ref{alg:cvpI}), where 
 $\vect{B}' = [(\vect{b}_1,0),\ldots, (\vect{b}_n,0),(\vect{t},\tau d/2)]$ is a basis for $\cL'$.
 Next, we run a number of iterations of the sieving Algorithm \ref{alg:multSieve} to obtain a number of vector pairs such that
 $\|\vect{y} \|_p \leq R = \frac{\xi d}{1-\gamma}+o(1)$.  
 Further details can be found in Algorithm \ref{alg:cvpI}.
 Note that in the algorithm, $\vect{v}|_{[n]}$ is the $n-$dimensional vector $\vect{v}'$ obtained by restricting $\vect{v}$
 to the first $n$ co-ordinates (with respect to the computational basis).
 
%-----------------------------------------------------------------------

\begin{algorithm}
 \caption{Approximate algorithm for $\cvpp$}
 \label{alg:cvpI}
 
 \KwIn{(i) A basis $\vect{B} = [\vect{b}_1, \ldots \vect{b}_n]$ of a lattice $L$, (ii) Target vector $\vect{t}$,
 (iii) Approximation factor $\tau$, (iv) $ 0 < \gamma <1 $, (v) $\xi$ such that $\frac{1}{2}\max(1,\tau/2) < \xi < 
 \frac{(1-\gamma)\tau}{2-\gamma} - c'$ where $c'$ is a small constant, 
 (vi)$\alpha >0$, (vii) $N \in \nat$ }
 \KwOut{A $2\tau-$approximate closest vector to $\vect{t}$ in $L$ }
 
 $d \leftarrow (1+\alpha) $ \;
 $T \leftarrow \emptyset; \qquad S'' \leftarrow \emptyset$ \;
 \While{$d \leq n\cdot \max_i \|\vect{b}_i\|_{p} $}
 {
    $S,S' \leftarrow \emptyset$ \;
    
    $\vect{B}' \rightarrow [(\vect{b}_1,0),\ldots, (\vect{b}_n,0),(\vect{t},\tau d/2)] $ 	\;
    $L' \rightarrow \lat(\vect{B}')$	\;
    $M \rightarrow \Span(\{(\vect{v},0):\vect{v} \in L \})$ \;
    
    \For{$i=1$ to $N$  \label{sample2:start}}
 {
    $\vect{e}_i \leftarrow_{\text{uniform}} \ballp_n(\vect{0},\xi\lambda) $ \; \label{sample2:v}
    $ \vect{y}_i \leftarrow \vect{e}_i \mod \fpar(\vect{B}) $ \;  \label{sample2:y}
   % $(\vect{e}_i,\vect{y}_i) \leftarrow \text{Sample}(\vect{B},\xi\lambda)$ 
   % using Algorithm \ref{alg:sample}   \label{multI:sample} \; 
    $S \leftarrow S \cup \{ (\vect{e}_i,\vect{y}_i) \}$	\; 
 }	\label{sample2:end}
    
    $R \leftarrow n \max_i \|\vect{b}_i\|_{p} $ \;
    
    \While{$R > \frac{\xi d}{1-\gamma}$ }
    {
      $S \leftarrow \text{sieve}(S,\gamma,R,\xi)$ using Algorithm \ref{alg:multSieve} \;
      $R \leftarrow \gamma R + \xi d$ 	\label{cvpI:R}	\; 
    }	\label{cvpI:sieveEnd}
    $S' \leftarrow \{ \vect{y}-\vect{e} : (\vect{e},\vect{y}) \in S \}$	\;
    Compute $\vect{w} \in S'$ such that $\|\vect{w}|_{[n]}\|_{p} = \min\{ \|\vect{v}'|_{[n]}\|_{p}:
    \vect{v}' \in S' \text{ and } (\vect{v}')_{n+1} \neq 0 \} $ \;
    $T \rightarrow T \cup \{ \vect{w} \} $	\; 
    $d \rightarrow d(1+\alpha)$ \;
 }

Let $\vect{v}_0$ be any vector in $T$ such that 
$\|\vect{v}_0|_{[n]}\|_{p}=\min\{ \|\vect{w}|_{[n]}\|_{p} : \vect{w} \in T \}$  \;
$\vect{v}_0' \leftarrow \vect{v}_0|_{[n]}$	\;
 \eIf{$(\vect{v}_0)_{n+1} = -\tau d/2$}
 {
    \Return $\vect{v}_0'+\vect{t}$	\;
 }
 {
    \Return $\vect{v}_0'-\vect{t}$	\;
 }
 
\end{algorithm}

%------------------------------------------------------------------------- 
 
 From Lemma \ref{lem:multI_finRad}, we have seen that after $\lceil \log_{\gamma} \Big( \frac{\xi}{nR_0(1-\gamma)} \Big)\rceil$
 iterations (where \linebreak$R_0 = n \cdot \max_i \|\vect{b}_i\|_p$), 
 $R \leq \frac{\xi \gamma}{n(1-\gamma)} + \frac{\xi d}{1-\gamma} \Big[ 1-\frac{\xi}{nR_0(1-\gamma)} \Big]$.
 %After repeating a few more iterations or with proper values of parameters $R$ can have the desired value.
 Thus, after the sieving iterations, the set $S'$ consists of vector pairs such that the corresponding lattice vector
 $\vect{v}$ has $\|\vect{v}\|_p \leq \frac{\xi d}{1-\gamma}+\xi d +c = \frac{\xi(2-\gamma)d}{1-\gamma}+o(1)$.
 
 Selecting $\xi < \frac{(1-\gamma)\tau}{2-\gamma} - o(1)$ ensures that our sieving algorithm does not return vectors from 
 $(\cL,0)-(k'\vect{t}, k'\tau d/2)$ for some $k'$ such that $|k'| \ge 2$.
 Then, every vector has  $\|\vect{v}\|_p < \tau d$, and so either $\vect{v} = \pm(\vect{z}'-\vect{t},0)$ or 
 $\vect{v} = \pm(\vect{z}-\vect{t},-\tau d/2)$ for some lattice vector $\vect{z},\vect{z}' \in \cL$.
 
 With similar arguments as in \cite{2018_AM2} (using the tossing argument outlined in \mbox{Section \ref{subsec:aks_lin})}, we can conclude
 that with some non-zero probability we have at least one vector in $\cL'\setminus (\cL \pm \vect{t},0)$ after the sieving 
 iterations. 

Thus, we obtain the following result.
\begin{Theorem}
 Let $\gamma \in (0,1)$, and for any $\tau > 1$ let $\xi > \max (1/2, \tau/4)$. 
  Given a full-rank lattice $\cL \subset \ratn^n$, there is a randomized algorithm that, for\\ 
  $\tau  = \frac{\xi (2-\gamma)}{1-\gamma} + o(1)$, approximates $\cvpp$ with a success probability of at 
  least $1/2$ and a space and time complexity of $2^{(\csp + \ccp) n + o(n)}$, where 
  $\ccp =\log \left(2+\frac{2}{\gamma} \right)$ and \\$\csp = - \log \left(0.5-\frac{1}{4\xi} \right)$. 
  %In particular, for $\gamma = 1/2 +o(1)$ and $\xi = \tau/3$, the algorithm runs in time 
  %$4^n \cdot \left(\frac{2\tau}{2\tau - 3}\right)^n$. 
 \label{thm:multI-approx-cvp}
\end{Theorem}
Again, using Mixed Sieve in Algorithm \ref{alg:multI}, we obtain space and time complexities of $2^{(\csp+\cmixp) n+o(n)}$ and $2^{(\csp+\cmixp+c_p)n+o(n)}$, respectively, where \\$\cmixp=\log \left(2+\frac{2}{A\gamma}\right)+c_p$ and $c_p=\log(1+A)$, respectively (in the Euclidean norm, the parameters are as described in Theorem \ref{thm:subQuad}).

%In $\ell_2$ norm using mixed sieving procedure we obtain time and space complexity of $2^{1.73n+o(n)}$ and a large constant approximation factor at parameters $\gamma=0.999$,$\xi=1$. In \cite{2011_LWXZ} the best running time reported is $2^{0.802n}$ for large approximation factor. 

%\input{./conclude}
{\section{Discussions}

In this paper, we have designed new sieving algorithms that work for any $\ell_p$ norm. A comparative performance evaluation has been given in Table \ref{tab:comp}. We achieve a better time complexity at the cost of space complexity for every $1\leq p\leq\infty$, except for the algorithm in \cite{2015_ADRS} that employs a Discrete Gaussian-based sieving algorithm and has better space and time complexity in the Euclidean norm. To the best of our knowledge, this algorithm does not work for any other norm.

\begin{table}[!htbp]
\centering 
\small
\begin{tabular}{|c|c|c|c|c|}
\hline
$p$ & Ref. & Type of sieve & Time complexity & Space complexity \\
\hline\hline
\multirow{2}{*}{$1\leq p\leq\infty$} & \cite{2009_BN} & Quadratic & $2^{3.849n+o(n)}$ & $2^{2.023n+o(n)}$  \\
\cline{2-5}
& This work & Linear & $2^{2.751n+o(n)}$  & $2^{2.751n+o(n)}$      \\
\hline\hline 
$p=\infty$ & \cite{2018_AM} & Linear & $2^{2.443n+o(n)}$  & $2^{2.443n+o(n)}$  \\
\hline\hline
\multirow{5}{*}{$p=2$} & \cite{2011_HPS} & Quadratic & $2^{2.571n+o(n)}$  & $2^{1.407n+o(n)}$ \\
\cline{2-5}
&This work & Linear & $2^{2.49n+o(n)}$ & $2^{2.49n+o(n)}$ \\
\cline{2-5} 
& \cite{2010_MV, 2011_HPS} & Quadratic & $2^{2.465n+o(n)}$ & $2^{1.233n+o(n)}$ \\
\cline{2-5}
& This work & Mixed & $2^{2.25n+o(n)}$ & $2^{2.25n+o(n)}$ \\
\cline{2-5}
& \cite{2015_ADRS} & DGS & $2^{n+o(n)}$ & $2^{n+o(n)}$ \\
\hline\hline
\end{tabular}
\caption{Comparison of the performance of various sieving algorithms in different $\ell_p$ norm. In the last row DGS stands fro Discrete Gaussian Sampling based sieve.}
\label{tab:comp}
\end{table}

\subsection{Future work}

An obvious direction for further research would be to design heuristic algorithms on these kind of sieving techniques and to study if these can be adapted to other computing environments like parallel computing.   

The major difference between our algorithm and the others like \cite{2001_AKS,2009_BN} is in the choice of the shape of the sub-regions in which we divide the ambient space (as has already been explained before). Due to this we get superior ``decodability'' in the sense that a vector can be efficiently mapped to a sub-region, at the cost of inferior space complexity, as described before. It might be interesting to study what other shapes of these sub-regions might be considered and what are the trade-offs we get.  

It might be possible to improve the bound on the number of hypercubes required to cover the hyperball. At least in the $\ell_{\infty}$ norm we have seen that the number of hypercubes may depend on the initial position of the smaller hypercube, whose translates cover the bigger hyperball. In fact it might be possible to get some lower bound on the complexity of this kind of approach.

\section*{Acknowledgement}

The author would like to acknowledge the anonymous reviewers for their helpful comments that have helped to improve the manuscript significantly. Research at IQC was supported in part by the Government of Canada through Innovation, Science and Economic Development Canada, Public Works and Government Services Canada and Canada First Research Excellence Fund.
\end{document}

%% file: preamble.tex
 \usepackage{fullpage}
\usepackage{graphicx}
\usepackage{upref}
\usepackage{enumerate}
\usepackage{latexsym}
\usepackage{pdfpages}
\usepackage[bookmarks]{hyperref}
\usepackage{color,graphics}
\usepackage{comment} 
 \usepackage{caption}
 \usepackage{subcaption}
 \usepackage{wrapfig}
 \usepackage{braket}
 \usepackage{multirow}
\usepackage{amssymb}
\usepackage{amsmath}
\usepackage{amsthm}
\usepackage{mathrsfs}
\usepackage{mathtools}

\usepackage{epsfig,graphicx}
\usepackage{algorithmic}
\usepackage[ruled,linesnumbered]{algorithm2e}
\usepackage{amsfonts}
\usepackage{tikz} 
\usepackage{varioref}
\usepackage[ansinew]{inputenc}
\usepackage{authblk}

\theoremstyle{plain}
\newtheorem{Theorem}{Theorem}[section]
\theoremstyle{plain}
\newtheorem{Lemma}{Lemma}[section]
\theoremstyle{plain}
\newtheorem{Corollary}{Corollary}[section]
\theoremstyle{plain}

\theoremstyle{plain}

\theoremstyle{plain}
\newtheorem{claim}{Claim}[section]

\theoremstyle{definition}
\newtheorem{Definition}{Definition}[section]

\theoremstyle{definition}
\newtheorem{fact}{Fact}[section]
\theoremstyle{definition}

\theoremstyle{remark}
\newtheorem{Remark}{Remark}[section]

\theoremstyle{definition}

\AtBeginDocument{}

\DeclareMathOperator{\real}{\mathbb{R}}
\DeclareMathOperator{\nat}{\mathbb{N}}

\newcommand{\ratn}{\mathbb{Q}}
\newcommand{\intg}{\mathbb{Z}}

\newcommand{\NP}{\text{NP}}

\newcommand{\RTIME}{\text{RTIME}}

\newcommand{\svp}{\textsf{SVP}}

\newcommand{\svpp}{\textsf{SVP}^{(p)}}

\newcommand{\cvp}{\textsf{CVP}}

\newcommand{\cvpp}{\textsf{CVP}^{(p)}}

\newcommand{\smp}{\textsf{SMP}}
\newcommand{\sivp}{\textsf{SIVP}}
\newcommand{\gsvp}{\textsf{GSVP}}

\newcommand{\vect}[1]{\mathbf{#1}}
\newcommand{\fpar}{\mathscr{P}}
\newcommand{\lat}{\mathscr{L}}
\newcommand{\bd}{\text{bd}}
\newcommand{\poly}{\text{poly}}

\newcommand{\ballp}{B^{(p)}}

\newcommand{\balli}{B^{(\infty)}}
\newcommand{\ballt}{B^{(2)}}

\newcommand{\Span}{\text{span}}
\newcommand{\minp}{\lambda^{(p)}}

\newcommand{\mini}{\lambda^{(\infty)}}
\newcommand{\mint}{\lambda^{(2)}}

\newcommand{\vol}{\text{vol}}

\newcommand{\bb}{\mathbf{b}}
\newcommand{\bB}{\mathbf{B}}
\newcommand{\bx}{\mathbf{x}}

\newcommand{\SVP}{\textsf{SVP}}
\newcommand{\SAP}{\textsf{SAP}}

\newcommand{\CVP}{\textsf{CVP}}
\newcommand{\R}{\mathbb{R}}
\newcommand{\cL}{\mathcal{L}}
\newcommand{\Z}{\mathbb{Z}}
\newcommand{\eps}{\varepsilon}
\newcommand{\cA}{\mathcal{A}}

\newcommand{\cc}{c_{c}^{(2)}}
\newcommand{\ccp}{c_c}
\newcommand{\cmixp}{c^{(p)}}
\newcommand{\cmix}{c^{(2)}}

\newcommand{\centre}{\mathcal{C}}
\newcommand{\centret}{\mathcal{C}^{(2)}}

\newcommand{\centrep}{\mathcal{C}}
\newcommand{\cb}{c_{b}^{(2)}}
\newcommand{\cbp}{c_b}

\newcommand{\cs}{c_s^{(2)}}
\newcommand{\csp}{c_s}

\newcommand{\cspace}{c_{space}^{(2)}}

\newcommand{\cspacep}{c_{space}}
\newcommand{\ctime}{c_{time}^{(2)}}

\newcommand{\ctimep}{c_{time}}

%% file: algoV3.bbl
\begin{thebibliography}{10}

\bibitem{1982_LLL}
Lenstra, A.K.; Lenstra, H.W., Jr.; Lov{\'a}sz, L.
\newblock Factoring polynomials with rational coefficients.
\newblock {\em Math. Ann.} \textbf{1982}, \emph{261}, 515--534.

\bibitem{1983_L}
Lenstra, H.W., Jr.
\newblock Integer programming with a fixed number of variables.
\newblock {\em Math. Oper. {R}es.} \textbf{1983}, \emph{8}, 538--548.

\bibitem{1987_K}
Kannan, R.
\newblock Minkowski's convex body theorem and integer programming.
\newblock {\em Math. Oper. Res.} \textbf{1987}, \emph{12}, 415--440.

\bibitem{2011_DPV}
Dadush, D.; Peikert, C.; Vempala, S.
\newblock Enumerative lattice algorithms in any norm via m-ellipsoid coverings.
\newblock In {Proceedings of the 2011 IEEE 52nd Annual Symposium on Foundations of Computer
  Science}, Palm Springs, CA, USA, 22--25 October 2011;  IEEE: Piscataway, NJ, USA, 2011; pp. 580--589.

\bibitem{2011_EHN}
Eisenbrand, F.; H{\"a}hnle, N.; Niemeier, M.
\newblock Covering cubes and the closest vector problem.
\newblock In { Proceedings of the Twenty-Seventh {A}nnual {S}ymposium on
  Computational {G}eometry}, Paris, France, 13--15 June 2011; ACM: New York, NY, USA, 2011; pp. 417--423.
  
\bibitem{1990_O}
Odlyzko, A.M.
\newblock The rise and fall of knapsack cryptosystems.
\newblock {\em Cryptol. Comput. Number Theory} \textbf{1990}, \emph{42}, 75--88.  

\bibitem{1998_JS}
Joux, A.; Stern, J.
\newblock Lattice reduction: A toolbox for the cryptanalyst.
\newblock {\em J. Cryptol.} \textbf{1998}, \emph{11}, 161--185.

\bibitem{2001_NS}
Nguyen, P.Q.; Stern, J.
\newblock The two faces of lattices in cryptology.
\newblock In {\em Cryptography and Lattices}; Springer:  Berlin/Heidelberg, Germany, %newly added information, please confirm  Reply : Confirmed
 2001; pp. 146--180.

\bibitem{1983_LM}
Landau, S.; Miller, G.L.
\newblock Solvability by radicals is in polynomial time.
\newblock In {Proceedings of the Fifteenth Annual ACM Symposium on Theory
  of {Computing}}%Please add address and specific time to the meeting （city, country, day month year）Reply : It has been added.
  ; ACM: New York, NY, USA, 1983; pp. 140--151.

\bibitem{1992_CJLOSS}
Coster, M.J.; Joux, A.; LaMacchia, B.A.; Odlyzko, A.M.; Schnorr, Cl.; Stern, J.
\newblock Improved low-density subset sum algorithms.
\newblock {\em Comput. Complex.} \textbf{1992}, \emph{2}, 111--128.

\bibitem{1996_A}
Ajtai, M.
\newblock Generating hard instances of lattice problems.
\newblock In {Proceedings of the Twenty-Eighth Annual ACM {S}ymposium on Theory of {C}omputing}, Philadelphia, PA, USA, 22--24 May 1996; ACM: New York, NY, USA, 1996; pp. 99--108.
  
\bibitem{2007_MR}
Micciancio, D.; Regev, O.
\newblock Worst-case to average-case reductions based on {G}aussian measures.
\newblock {\em SIAM J. Comput.} \textbf{2007}, \emph{37}, 267--302.  

\bibitem{2009_G}
Gentry, C.
\newblock Fully homomorphic encryption using ideal lattices.
\newblock In {Proceedings of the S{TOC}'09---{P}roceedings of the 2009 {ACM} {I}nternational
  {S}ymposium on {T}heory of {C}omputing}, Bethesda, MD, USA, 31 May--2 June 2009; ACM: New York, NY, USA, 2009;  pp. 169--178.
 
  
\bibitem{2009_R}
Regev, O.
\newblock On lattices, learning with errors, random linear codes, and
  cryptography.
\newblock {\em J. ACM} \textbf{2009}, \emph{56}, 34.%, \hl{40}%Is this necessary? Can this be deleted?  Reply : Yes. Removed.
.  

\bibitem{2011_BV}
Brakerski, Z.; Vaikuntanathan, V.
\newblock Efficient fully homomorphic encryption from (standard) {LWE}.
\newblock In {Proceedings of the FOCS}, Palm Springs, CA, USA, 23--25 October 2011;  IEEE: Piscataway, NJ, USA,
2011; pp. 97--106.

\bibitem{2013_BLPRS}
Brakerski, Z.; Langlois, A.; Peikert, C.; Regev, O.; Stehl{\'e}, D.
\newblock Classical hardness of learning with errors.
\newblock In {Proceedings of the STOC}, Palo Alto, CA, USA, 1--4 June, 2013; pp. 575--584.

\bibitem{2014_BV}
Brakerski, Z.; Vaikuntanathan, V.
\newblock Lattice-based {FHE} as secure as {PKE}.
\newblock In {Proceedings of the ITCS},  Princeton, NJ, USA, 12--14 January 2014; pp. 1--12.

\bibitem{Pei16}
Peikert, C.
\newblock A decade of lattice cryptography.
\newblock {\em Found. Trends Theor. Comput. Sci.} \textbf{2016}, \emph{10}, 283--424.
  
\bibitem{2017_DLLSSS}
Ducas, L.; Lepoint, T.; Lyubashevsky, V.; Schwabe, P.; Seiler, G.; Stehl{\'e}, D.
\newblock \emph{Crystals--Dilithium: Digital Signatures from Module Lattices};
\newblock \emph{IACR Transactions on Symmetric Cryptology}, \textbf{2018}, 238--268.
%\newblock Technical Report ; \hl{IACR Cryptology ePrint Archive}%Please add the location of the publisher.  Reply : Updated the reference.
%: 2017; Volume 633, p. 2017.

\bibitem{2015_ADRS}
Aggarwal, D.; Dadush, D.; Regev, O.; Stephens{-}Davidowitz, N.
\newblock Solving the {S}hortest {V}ector {P}roblem in $2^n$ time via {D}iscrete
  {G}aussian sampling.
\newblock {In Proceedings of the STOC}, Portland, OR, USA, 14--17 June%MDPI: Newly added information, please confirm.
2015.
%\newblock Available online: \url{http://arxiv.org/abs/1412.7994} (\hl{accessed on}%Please add accessed date.).  Reply : We do not need this.

\bibitem{2001_AKS}
Ajtai, M.; Kumar, R.; Sivakumar, D.
\newblock A sieve algorithm for the shortest lattice vector problem.
\newblock {In Proceedings of the  STOC, Heraklion, Greece, 6--8 July} 2001; pp. 601--610.

\bibitem{1985_FP}
Fincke, U.; Pohst, M.
\newblock Improved methods for calculating vectors of short length in a
  lattice, including a complexity analysis.
\newblock {\em Math. Comput.} \textbf{1985}, \emph{44}, 463--471.

\bibitem{1987_S}
Schnorr, C.-P.
\newblock A hierarchy of polynomial time lattice basis reduction algorithms.
\newblock {\em Theor. Comput. {S}ci.} \textbf{1987}, \emph{53}, 201--224.

\bibitem{2013_MV}
Micciancio, D.; Voulgaris, P.
\newblock A deterministic single exponential time algorithm for most lattice
  problems based on {V}oronoi cell computations.
\newblock {\em SIAM J. Comput.} \textbf{2013}, \emph{42}, 1364--1391.

\bibitem{2013_DV}
Dadush, D.; Vempala, S.S.
\newblock Near-optimal deterministic algorithms for volume computation via
  m-ellipsoids.
\newblock {\em Proc. Natl. Acad. Sci. USA} \textbf{2013},
  \emph{110}, 19237--19245.

\bibitem{2011_HPS}
Hanrot, G.; Pujol, X.; Stehl{\'e}, D.
\newblock Algorithms for the shortest and closest lattice vector problems.
\newblock In {\em International Conference on Coding and Cryptology}; Springer:  Berlin/Heidelberg, Germany, %newly added information, please confirm  Reply : It is correct
2011; pp. 159--190.

\bibitem{2010_MV}
Micciancio, D.; Voulgaris, P.
\newblock Faster exponential time algorithms for the shortest vector problem.
\newblock In {Proceedings of the SODA}, Austin, TX, USA, 17--19 January 2010; pp. 1468--1480.

\bibitem{2009_PS}
Pujol, X.; Stehl{\'e}, D.
\newblock Solving the shortest lattice vector problem in time $2^{2.465n}$.
\newblock {\em IACR Cryptol. ePrint Arch.} \textbf{2009}, \emph{2009}, 605.

\bibitem{2017_AS}
Aggarwal, D.; Stephens-Davidowitz, N.
\newblock Just take the average! an embarrassingly simple $2^{n} $-time
  algorithm for {SVP} (and {CVP}).
\newblock {\em arXiv} \textbf{2017}, arXiv:1709.01535.

\bibitem{2011_LWXZ}
Liu, M.; Wang, X.; Xu, G.; Zheng, X.
\newblock Shortest lattice vectors in the presence of gaps.
\newblock {\em IACR Cryptol. Eprint Arch.} \textbf{2011}, \emph{2011}, 139.

\bibitem{2008_NV}
Nguyen, P.Q.; Vidick, T.
\newblock Sieve algorithms for the shortest vector problem are practical.
\newblock {\em J. Math. Cryptol.} \textbf{2008}, \emph{2}, 181--207.

\bibitem{2011_WLTB}
Wang, X.; Liu, M.; Tian, C.; Bi, J.
\newblock Improved {N}guyen-{V}idick heuristic sieve algorithm for shortest vector problem.
\newblock In {Proceedings of the AsiaCCS}, Hong Kong, China, 22--24 March 2011; pp. 1--9.

\bibitem{2016_BDGL}
Becker, A.; Ducas, L.; Gama, N.; Laarhoven, T.
\newblock New directions in nearest neighbor searching with applications to
  lattice sieving.
\newblock In {Proceedings of the Twenty-Seventh Annual ACM-SIAM Symposium
  on Discrete Algorithms}, Arlington, VA, USA, 10--12 January 2016; Society for Industrial and Applied
  Mathematics: Philadelphia, PA, USA, 2016; pp. 10--24.

\bibitem{2015_LW}
Laarhoven, T.; de Weger, B.
\newblock Faster sieving for shortest lattice vectors using spherical
  locality-sensitive hashing.
\newblock In {\em International Conference on Cryptology and Information
  Security in Latin America}; Springer:  Berlin/Heidelberg, Germany,
 2015; pp. 101--118.  
  
\bibitem{2016_BL}
Becker, A.; Laarhoven, T.
\newblock Efficient (ideal) lattice sieving using cross-polytope {LSH}.
\newblock In {\em International Conference on Cryptology in Africa}; Springer:  Berlin/Heidelberg, Germany,
 2016; pp.
  3--23.  
  
\bibitem{2017_HK}
Herold, G.; Kirshanova, E.
\newblock Improved algorithms for the approximate k-list problem in {E}uclidean
  norm.
\newblock In {\em IACR International Workshop on Public Key Cryptography};
   Springer:  Berlin/Heidelberg, Germany,
 2017; pp. 16--40.
  
\bibitem{2018_HKL}
Herold, G.; Kirshanova, E.; Laarhoven, T.
\newblock Speed-ups and time--memory trade-offs for tuple lattice sieving.
\newblock In {\em IACR International Workshop on Public Key Cryptography};
  Springer:  Berlin/Heidelberg, Germany, 
 2018; pp. 407--436.
  
\bibitem{2018_LM}
Laarhoven, T.; Mariano, A.
\newblock Progressive lattice sieving.
\newblock In {\em International Conference on Post-Quantum Cryptography}; Springer: Berlin/Heidelberg, Germany, 
   2018; pp.
  292--311.
  
\bibitem{2016_MB}
Mariano, A.; Bischof, C.
\newblock Enhancing the scalability and memory usage of hash sieve on multi-core
  {CPU}s.
\newblock In {Proceedings of the 2016 24th Euromicro International Conference on Parallel,
  Distributed, and Network-Based Processing (PDP)}, Heraklion, Greece, 17--19 February 2016; IEEE: Piscataway, NJ, USA, 2016; pp. 545--552.
  
\bibitem{2017_MLB}
Mariano, A.; Laarhoven, T.; Bischof, C.
\newblock A parallel variant of {LD} sieve for the {SVP} on lattices.
\newblock In {Proceedings of the 2017 25th Euromicro International Conference on Parallel,
  Distributed and Network-based Processing (PDP)}, St. Petersburg, Russia, 6--8 March 2017; IEEE: Piscataway, NJ, USA, 2017; pp. 23--30.
  

\bibitem{2017_YKYC}
Yang, S.-Y.; Kuo, P.-C.; Yang, B.-Y.; Cheng, C.-M.
\newblock Gauss sieve algorithm on {GPU}s.
\newblock In {\em Cryptographers' Track at the RSA Conference}; 
  Springer:  Berlin/Heidelberg, Germany, 
 2017; pp. 39--57.  
  
\bibitem{2018_D}
Ducas, L.
\newblock Shortest vector from lattice sieving: A few dimensions for free.
\newblock In {\em Annual International Conference on the Theory and
  Applications of Cryptographic Techniques};  Springer:  Berlin/Heidelberg, Germany,
 2018; pp. 125--145.

\bibitem{2019_ADHKPS}
Albrecht, M.R.; Ducas, L.; Herold, G.; Kirshanova, E.; Postlethwaite, E.W.; Stevens, M.
\newblock The general sieve kernel and new records in lattice reduction.
\newblock In {\em Annual International Conference on the Theory and
  Applications of Cryptographic Techniques}; Springer: Berlin/Heidelberg, Germany, 
   2019; pp. 717--746.  
   
\bibitem{1999_GMSS}
Goldreich, O.; Micciancio, D.; Safra, S.; Seifert, J.-P.
\newblock Approximating shortest lattice vectors is not harder than
  approximating closest lattice vectors.
\newblock {\em Inf. Process. Lett.} \textbf{1999}, \emph{71}, 55 -- 61.

\bibitem{2002_AKS}
Ajtai, M.; Kumar, R.; Sivakumar, D.
\newblock Sampling short lattice vectors and the closest lattice vector
  problem.
\newblock In Proceedings of the {CCC} Beijing, China, 15--20 April 2002; pp. 41--45.

\bibitem{2015_ADS}
Aggarwal, D.; Dadush, D.; Stephens-Davidowitz, N.
\newblock Solving the closest vector problem in $2^n$ time--the {D}iscrete {G}aussian strikes again!
\newblock In {Proceedings of the Foundations of Computer Science (FOCS), 2015 IEEE 56th Annual
  Symposium}, Berkeley, CA, USA, 18--20 October 2015; IEEE: Piscataway, NJ, USA, 2015; pp. 563--582.  

\bibitem{2009_BN}
Bl{\"o}mer, J.; {Naewe, S.} %Please confirm whether ref 4 and 66 are duplicates Reply : They are by the same author, published in two different places, one in journal and another in conference. It is better to keep both.
\newblock Sampling methods for shortest vectors, closest vectors and successive
  minima.
\newblock {\em Theor. Comput. Sci.} \textbf{2009}, \emph{410}, 1648--1665.

\bibitem{2008_AJ}
Arvind, V.; Joglekar, P.S.
\newblock Some sieving algorithms for lattice problems.
\newblock In {\em LIPIcs-Leibniz International Proceedings in Informatics};
  Schloss Dagstuhl-Leibniz-Zentrum f{\"u}r Informatik: Wadern, Germany, 2008; Volume~2.
  
\bibitem{2018_AM}
{Aggarwal, D.; Mukhopadhyay, P.} %Please confirm whether ref 50 and 62 are duplicates  Reply : They are the different versions of the same paper. There are some more results in 62. So it is better to keep both.
\newblock Improved algorithms for the shortest vector problem and the closest
  vector problem in the infinity norm.
\newblock In {Proceedings of the 29th International Symposium on Algorithms and Computation
  (ISAAC 2018)}, {Yilan County, Taiwan, 17--19 December 2018}; Schloss Dagstuhl-Leibniz-Zentrum fuer Informatik: {Wadern, Germany}, 2018.  
  
\bibitem{1981_vE}
van Emde~Boas, P.
\newblock \emph{Another {NP}-Complete Partition Problem and the Complexity of Computing Short Vectors in a Lattice};
\newblock Technical Report; Department of Mathematics, University of Amsterdam, ~{1981}%Please add the publisher and location. Reply : Added whatever information was available.
.  

\bibitem{1998_A}
Ajtai, M.
\newblock The shortest vector problem in $\ell_2$ is {NP}-hard for randomized reductions.
\newblock In {Proceedings of the Thirtieth Annual ACM Symposium on Theory
  of {C}omputing}, Dallas, TX, USA, 24--26 May 1998; ACM: New York, NY, USA, 1998; pp. 10--19.

\bibitem{2001_M}
Micciancio, D.
\newblock The shortest vector problem is {NP}-hard to approximate to within
  some constant.
\newblock {\em SIAM J. Comput.} \textbf{2001}, \emph{30}, 2008--{2035}.%We removed extra information, please confirm.  Reply : Confirmed.

  
\bibitem{2003_DKRS}
Dinur, I.; Kindler, G.; Raz, R.; Safra, S.
\newblock Approximating {CVP} to within almost-polynomial factors is {NP}-hard.
\newblock {\em Combinatorica} \textbf{2003}, \emph{23}, 205--243.  

\bibitem{2000_D}
Dinur, I.
\newblock Approximating {$SVP_{\infty}$} to within almost-polynomial factors is {NP}-hard.
\newblock In {\em Italian Conference on Algorithms and Complexity}; Springer:  Berlin/Heidelberg, Germany,
 2000; pp.
  263--276.

\bibitem{2019_M}
Mukhopadhyay, P.
\newblock The projection games conjecture and the hardness of approximation of
  SSAT and related problems.
\newblock {\em J. Comput. Syst. Sci.} \textbf{2021}, \emph{123}, 186--201.  
  
\bibitem{2015_M}
Moshkovitz, D.
\newblock The projection games conjecture and the {NP}-hardness of ln $n$-approximating Set-Cover.
\newblock {\em Theory Comput.} \textbf{2015}, \emph{11}, 221--235.

\bibitem{2005_K}
Khot, S.
\newblock Hardness of approximating the shortest vector problem in lattices.
\newblock {\em J. ACM} \textbf{2005}, \emph{52}, 789--808.


\bibitem{2012_HR}
Haviv, I.; Regev, O.
\newblock Tensor-based hardness of the shortest vector problem to within almost
  polynomial factors.
\newblock {\em Theory Comput.} \textbf{2012}, \emph{8}, 513--531.


\bibitem{2017_BGS}
Bennett, H.; Golovnev, A.; Stephens-Davidowitz, N.
\newblock On the quantitative hardness of {CVP}.
\newblock {\em arXiv} \textbf{2017}, arXiv:1704.03928.

\bibitem{2018_AsD}
Aggarwal, D.; Stephens-Davidowitz, N.
\newblock (gap/S){ETH} hardness of {SVP}.
\newblock In {Proceedings of the 50th Annual ACM SIGACT Symposium on Theory
  of Computing}, Los Angeles, CA, USA, 25--29 June 2018; ACM: New York, NY, USA, 2018; pp. 228--238.
  
\bibitem{2018_AM2}
Aggarwal, D.; Mukhopadhyay, P.
\newblock Improved algorithms for the shortest vector problem and the closest
  vector problem in the infinity norm.
\newblock {\em arXiv} \textbf{2018}, arXiv:1801.02358.

\bibitem{2013_DK}
Dadush, D.; Kun, G.
\newblock Lattice sparsification and the approximate closest vector problem.
\newblock In {Proceedings of the Twenty-Fourth annual ACM-SIAM Symposium on
  Discrete Algorithms}, New Orleans LA, USA, 6--8 January 2013; Society for Industrial and Applied
  Mathematics: Philadelphia, PA, USA, 2013; pp. 1088--1102.
  
\bibitem{1991_DFK}
Dyer, M.; Frieze, A.; Kannan, R.
\newblock A random polynomial-time algorithm for approximating the volume of
  convex bodies.
\newblock {\em J. ACM JACM} \textbf{1991}, \emph{38}, 1--17.

\bibitem{2000_GG}
Goldreich, O.; Goldwasser, S.
\newblock On the limits of nonapproximability of lattice problems.
\newblock {\em J. Comput. Syst. Sci.} \textbf{2000}, \emph{60}, 540--563.  
  
\bibitem{2007_BN}
Bl{\"o}mer, J.; Naewe, S.
\newblock Sampling methods for shortest vectors, closest vectors and successive
  minima.
\newblock In {\em International Colloquium on Automata, Languages, and
  Programming}; Springer:  Berlin/Heidelberg, Germany, 
 2007; pp. 65--77.

\bibitem{1978_KL}
Kabatiansky, G.A.; Levenshtein, V.I.
\newblock On bounds for packings on a sphere and in space.
\newblock {\em Probl. Peredachi Informatsii} \textbf{1978}, \emph{14}, 3--25.

\bibitem{1999_P}
Pisier, G.
\newblock {\em The Volume of Convex Bodies and Banach Space Geometry};
\newblock Cambridge University Press: Cambridge, UK, 1999; Volume~94.

\bibitem{2009_R1}
Regev, O.
\newblock \emph{{Lecture Notes on Lattices in Computer Science}}, New York University, U.S.%Please add the publisher and location.  Reply : Added.
;
\newblock 2009.  

\bibitem{1986_B}
Babai, L.
\newblock On {L}ov\'asz' lattice reduction and the nearest lattice point
  problem.
\newblock {\em Combinatorica} \textbf{1986}, \emph{6}, 1--13.


\end{thebibliography}

\begin{thebibliography}{999}
% Reference 1
\bibitem[Author1(year)]{ref-journal}
Author~1, T. The title of the cited article. {\em Journal Abbreviation} {\bf 2008}, {\em 10}, 142--149.
% Reference 2
\bibitem[Author2(year)]{ref-book1}
Author~2, L. The title of the cited contribution. In {\em The Book Title}; Editor1, F., Editor2, A., Eds.; Publishing House: City, Country, 2007; pp. 32--58.
% Reference 3
\bibitem[Author3(year)]{ref-book2}
Author 1, A.; Author 2, B. \textit{Book Title}, 3rd ed.; Publisher: Publisher Location, Country, 2008; pp. 154--196.
% Reference 4
\bibitem[Author4(year)]{ref-unpublish}
Author 1, A.B.; Author 2, C. Title of Unpublished Work. \textit{Abbreviated Journal Name} stage of publication (under review; accepted; in~press).
% Reference 5
\bibitem[Author5(year)]{ref-communication}
Author 1, A.B. (University, City, State, Country); Author 2, C. (Institute, City, State, Country). Personal communication, 2012.
% Reference 6
\bibitem[Author6(year)]{ref-proceeding}
Author 1, A.B.; Author 2, C.D.; Author 3, E.F. Title of Presentation. In Title of the Collected Work (if available), Proceedings of the Name of the Conference, Location of Conference, Country, Date of Conference; Editor 1, Editor 2, Eds. (if available); Publisher: City, Country, Year (if available); Abstract Number (optional), Pagination (optional).
% Reference 7
\bibitem[Author7(year)]{ref-thesis}
Author 1, A.B. Title of Thesis. Level of Thesis, Degree-Granting University, Location of University, Date of Completion.
% Reference 8
\bibitem[Author8(year)]{ref-url}
Title of Site. Available online: URL (accessed on Day Month Year).
\end{thebibliography}
